\documentclass[traditabstract]{aa}

\usepackage{graphicx}
\usepackage{txfonts}
\usepackage{longtable,lscape}
\usepackage{xcolor}

\usepackage{natbib}
\bibpunct{(}{)}{;}{a}{}{,} 

\usepackage{gensymb}

\newcommand{\planck}{\textit{Planck }}
\newcommand{\Planck}{\textit{Planck }}

\usepackage{multirow}

\begin{document}

\title{Velocity dispersion and dynamical masses for 388 Galaxy Clusters and groups. Calibrating the $M_{\rm SZ}$--$M_{\rm dyn}$ scaling relation for the PSZ2 sample}

\subtitle{}

\author{A.~Aguado-Barahona \inst{1,2} \and J.A.~Rubiño-Martín \inst{1,2}\and  A.~Ferragamo \inst{1,2} \and  R.~Barrena \inst{1,2} \and A.~Streblyanska   \inst{1,2} \and D.~Tramonte \inst{3,1,2}}

\institute{Instituto de Astrofísica de Canarias, C/Vía Láctea s/n, E-38205 La Laguna, Tenerife, Spain\\
              \email{alejandro.aguado.barahona@gmail.com,aaguado@iac.es}
         \and
             Universidad de La Laguna, Departamento de Astrofísica, E-38206
                                La Laguna, Tenerife, Spain
         \and
            Purple Mountain Observatory, No. 8 Yuanhua Road, Qixia District, Nanjing 210034, China \\}

   \date{}

\abstract
{The second catalogue of \Planck Sunyaev-Zeldovich (SZ) sources, hereafter PSZ2, represents the largest galaxy cluster sample selected by means of their SZ signature in a full-sky survey.
Using telescopes at the Canary Island observatories, we conducted the long-term observational program 128- MULTIPLE-16/15B (hereafter LP15), a large and complete optical follow-up campaign of all the unidentified PSZ2 sources in the northern sky, with declinations above $-15^\circ$ and no correspondence in the first \Planck catalogue PSZ1. 
This paper is the third and last in the series of LP15 results, after \cite{paper4} and \cite{paper5}, and presents all the spectroscopic observations of the full program. 

We complement these LP15 spectroscopic results with Sloan Digital Sky Survey (SDSS) archival data and other observations from a previous program (ITP13-08), and present a catalog of 388 clusters and groups of galaxies including estimates of their velocity dispersion. The majority of them (356) are the optical counterpart of a PSZ2 source. 
A subset of 297 of those clusters is used to construct the $M_{\rm SZ}-M_{\rm dyn}$ scaling relation, based on the estimated SZ mass from \Planck measurements and our dynamical mass estimates. We discuss and correct for different statistical and physical biases in the estimation of the masses, such as the Eddington bias when estimating $M_{SZ}$ and the aperture and the number of galaxies used to calculate $M_{dyn}$.
The SZ-to-dynamical mass ratio for those 297 PSZ2 clusters is $(1-B) = 0.80\pm0.04$ (stat) $\pm 0.05$ (sys), with only marginal evidence for a possible mass dependence of this factor. 
Our value is consistent with previous results in the literature, but presents a significantly smaller uncertainty due to the use of the largest sample size for this type of studies. 
 }

\titlerunning{LP15 spectroscopy and mass scaling relation in PSZ2}
\maketitle
%

\section{Introduction}
        
Galaxy Clusters (GCs) are the most massive bound objects in the Universe that emerge from the hierarchical structure formation \citep{Peebles80}. They are excellent tracers of the matter density distribution on the relevant scales for cosmological studies. Indeed, the evolution of the GC abundance with mass and redshift is very sensitive to the amplitude of the matter density fluctuations $\sigma_8$ and the mean matter density of the Universe $\Omega_m$ \citep{Allen11}.

The ESA’s \planck\footnote{\Planck\ \url{http://www.esa.int/Planck} is a project of the European Space Agency (ESA) with instruments provided by two scientific consortia funded by ESA member states and led by Principal Investigators from France and Italy, telescope reflectors provided through a collaboration between ESA and a scientific consortium led and funded by Denmark, and  additional contributions from NASA (USA).} mission \citep{Planck14I} provided, for the first time, the possibility to detect galaxy clusters using their 
Sunyaev-Zel'dovich \citep[hereafter SZ,][]{SZ72} signature in a full sky survey. The associated \Planck\ data products have been widely used to study mass scaling relations. In particular, this paper is based on PSZ2 \citep{PSZ2}, the second \planck\ catalog of SZ sources derived from the full 29 months mission data. This catalog is built with the combined results from three cluster detection codes (MMF1, MMF3 and PwS), as described in detail in \cite{PSZ1,PSZ2}. 

The PSZ2 catalog contains 1653 detections, and was partially validated at the time of publication using external X-ray, optical, near-IR and SZ data \citep{PSZ2}. This first validation process began with a cross-match with the PSZ1 \citep{PSZ1}. After that, the search for possible counterparts continued in X-rays with the MCXC catalog \citep{Piffaretti11}, which is based on the \textit{ROSAT} All Sky Survey \citep[RASS,][]{Voges99,Voges00}, and the serendipitous \textit{ROSAT} and Einstein cluster catalogs. In the optical and near-IR, they used the Sloan Digital Sky Survey \citep[SDSS][]{York00}, the redMaPPer catalog \citep{Rykoff14}, and the AllWISE mid-infrared source catalog \citep{Cutri13}. Finally, they also used SZ information, such as the catalogs obtained by the South Pole Telescope \citep[SPT,][]{Bleem15}, by the Atacama Cosmology Telescope \citep[ACT,][]{Hasselfield13} and by direct follow-up with the Arc-minute Micro-kelvin Interferometer \citep[AMI,][]{Perrott15}.

This paper is the third (and last) in the series of publications associated with the observational program {\tt 128-MULTIPLE-16/15B} (hereafter {\tt LP15}), an optical follow-up campaign of all the unidentified (at the time of publication) 190 PSZ2 sources in the northern sky, with declinations above $\delta = -15^\circ$ and no correspondence in the first \planck\ catalog PSZ1. Papers I \citep{paper4} and II \citep{paper5} in this series already presented the full program, the imaging results and the full validation analysis of the LP15 sample, including the confirmation of new GCs and their corresponding redshifts. In this paper III, we present all the spectroscopic observations of the program, including velocity dispersion and dynamical mass estimates in some cases.
These LP15 observations are complemented here with the use of SDSS archival data, allowing us to increase significantly (259 new objects) the number of PSZ2 clusters with spectroscopic information in the northern sky. 
This work also makes use of the validation papers \cite{paper1}, \cite{paper2} and \cite{paper3}, associated with the study of the PSZ1 catalog by means of the observational program {\tt ITP13-08}. 

The mass of a GC is not directly measurable. It is an unfortunate fact that can be circumvented using scaling relations based on different mass proxies \citep{Pratt19}. X-ray mass measurements are based on the assumption of hydrostatic equilibrium, and use preferably the product of gas mass and temperature ($Y_X = kTM_{gas}$) due to the low scatter of this quantity \citep[e.g. ][]{Krastov06}. The SZ effect can also be used to estimate masses. The usual proxy in this case is the spherically integrated Comptonization parameter, $Y_{SZ}$, which is related to the integrated electron pressure along the line-of-sight. 
Optical/dynamical mass methods are based on the assumption of dynamical equilibrium where the galaxies are the main ingredients. They use the velocity dispersion as a mass proxy, via the virial theorem. This mass estimation is often biased, due to violations of the hydrostatical or dynamical equilibrium, the temperature structure \citep{Rasia14}, selection and/or observational effects. To account for these deviations for all these methods, the mass bias parameter $(1-b)$ is introduced as $M_X = (1-b)_X M_{true}$ for $X$ = X-ray, SZ, dynamical masses, respectively. One of the main aims of this work is to characterize this mass bias for the PSZ2 sample in the case of dynamical masses, in order to obtain unbiased SZ masses which could be used for cosmological studies. The estimation of the mass bias has been studied recently by a large number of groups in the community: \cite{Ruel14}, \cite{Sifon16} and \cite{Amodeo18} using SZ and dynamical masses; or \cite{vonderLinden14}, \cite{Hoekstra15}, \cite{Smith16}, \cite{Battaglia16}, \cite{Sereno17}, \cite{PennaLima17}, \cite{Medezinski18} and \cite{Miyatake19} using SZ and weak-lensing masses.

The estimation of the mass bias parameter is not a straight-forward task. Although there are methods that give accurate estimations of the mass of an individual cluster, they usually come with large statistical or systematic errors \citep[see e.g.][]{Tremaine02, Kelly07, deMartino16, Sereno17}. This fact combined with the existence of intrinsic scatter in the fitted relations, make this estimation a complex task.
Thus, the appropriate characterization of the selected regression method used in each case is mandatory. In Appendix~\ref{app:B} we address this topic of linear regression with errors in both axes and intrinsic scatter, for the particular case of our sample.

This paper is structured as follows. Section~\ref{sec:sample} describes our reference sample and the corresponding data sets, including the final results of the LP15 program. Section~\ref{sec:vd} illustrates our methodology for the velocity dispersion estimates. In Section~\ref{sec:mass}, we present our dynamical mass estimates, and compare them to the SZ masses. Section~\ref{sec:sr} shows the results for the characterization of the scaling relation $M_{SZ}-M_{dyn}$ in the PSZ2 north, and the results for the mass bias factor $(1-b)$. Section~\ref{sec:con} presents the conclusions. Appendix~\ref{app:A} provides the results for the 362 galaxy clusters and groups in table format. Appendix~\ref{app:B} describes the simulations performed to validate the various regression methods.
Throughout this paper, we adopt a $\Lambda$CDM cosmology with $\ohm_{\rm m} = 0.3075$, $\ohm_{\Lambda} = 0.691,$ and H$_0 = 67.74$\,km\,s$^{-1}$\,Mpc$^{-1}$  \citep{2016A&A...594A..13P}.
        

\section{The reference sample}
\label{sec:sample}

The PSZ2 catalog \citep{PSZ2} is the largest full sky sample of GCs detected via the SZ effect. It consists of 1653 detections. At the time of its publication, 1203 sources were confirmed as counterparts in other wavelengths. After a great effort by the community, 1425 objects in total have been validated to date \citep{2018AstL...44..297B,2019ApJ...871..188B,paper4,paper5}. 
From here, the PSZ2-North sub-sample is defined as in papers I and II \citep{paper4,paper5}, as those 1003 objects within the PSZ2 catalog with $Dec.>-15\degree$. 

This article presents the velocity dispersion and dynamical mass for a sample of 388 objects (see Table~\ref{tab:sample}). The majority of them (356) are the optical counterpart of a SZ source in the PSZ2-North sample (note that double detections are counted as two different clusters). Six clusters are the optical counterpart of a PSZ2 source but outside the PSZ2-North sample. The remaining 26 objects were found during the process of analysis of the fields in which a SZ source is present, but were not associated with the SZ signal (see Sect.~\ref{sec:bey_psz2N} for details). Each object in our sample comes from one particular data set. These data sets are described in the following sub-sections.

\begin{table*}
\centering
\caption{Summary of the data sets.}
\begin{tabular}{c c c c|c c}
\hline \hline
\noalign{\smallskip}
Data set    & PSZ2-North  & Others in PSZ2 & Beyond PSZ2 & 
\multicolumn{2}{c}{Scaling relation}\\
 & & (see Sect.~\ref{sec:others_psz2}) & (see Sect.~\ref{sec:bey_psz2N}) & $1.5\times r_{200}$ & $1\times r_{200}$     \\
\hline
\noalign{\smallskip}
LP15        &  63 & 6 &  13 &  48 &  44  \\
ITP13       &  43 & 0 &   4 &  38 &  33  \\
SDSS        & 250 & 0 &   9 & 211 & 184  \\
\hline
Total       & 356 & 6 &  26 & 297 & 261  \\
\hline \hline
\end{tabular}
\label{tab:sample}
\end{table*}

The possible presence of interlopers inside the cluster radial velocity catalogs might bias the velocity dispersion and mass estimates \citep{Mamon11}. For this reason, we decided to analyze the clusters in two ways, trying to characterize the presence of this source of error. During the member selection process, explained in detail in Sect.~\ref{sec:vd}, we use two different apertures, namely 1 and 1.5 $\times r_{200}$, to select the cluster members. The comparison of the mass bias in both samples gives no significant difference between them, so we can safely assume that the number of interlopers within 1 and 1.5 $\times r_{200}$ is sufficiently small (compared to our statistical error) to not account for them. Table~\ref{tab:sample} includes the details about these sub-samples.

\begin{figure}
\centering \includegraphics[width=\columnwidth]{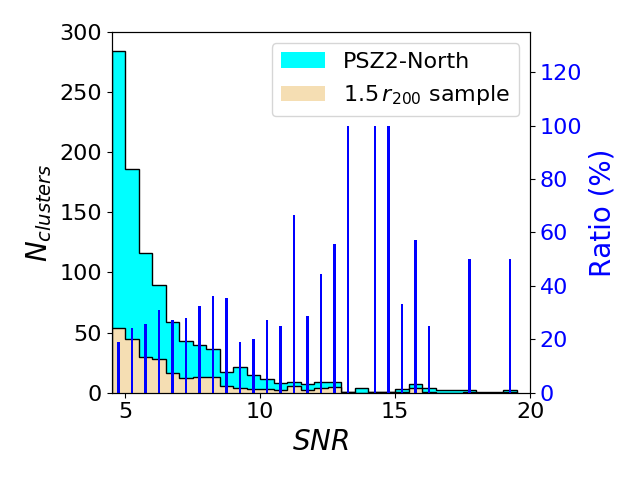}
\caption{PSZ2 cluster counts as a function of the signal-to-noise ratio (SNR) of the SZ detection. The PSZ2-North sample is represented in light blue, and the $1.5 \times r_{200}$ sample is represented in wheat. Dark blue bars represent the ratio between the $1.5 \times r_{200}$ sample and the total number of clusters in the PSZ2-North sample. The bin size is $0.5$. } 
\label{fig:histSNR}
\end{figure}

Figure~\ref{fig:histSNR} shows the number of clusters inside the $1.5 \times r_{200}$ sub-sample in comparison with the total number of objects in the PSZ2-North sample, as a function of the signal-to-noise ratio (SNR) in the PSZ2 catalog. Our sample covers the full range of SNR values, being approximately $30$\,\% of the total PSZ2-North sample. This fact allows us to consider this sample as statistically representative to infer global properties of the full PSZ2-North sample.

\subsection{LP15 data set}
\label{sec:lp15}

The {\tt 128-MULTIPLE-16/15B} follow-up program {\tt LP15} was designed to observe all PSZ2-North sources with no confirmed counterparts at the moment of the catalog's publication. 
This original LP15 sample contains 190 objects \citep{paper5}. The program had two main goals: to validate the SZ sources by finding their optical counterparts, and to use them for the calibration of the $M_{SZ}-M_{dyn}$ scaling relation. The validation process was published in \cite{paper4} and \cite{paper5}. In total, 184 sources were observed, being 81 of them confirmed as optical counterparts of the PSZ2 detections.

The {\tt LP15} program was performed during four consecutive semesters (2015B, 2016A, 2016B and 2017A). Due to technical telescope issues we were not able to complete this program in time, so we were granted with other four observing nights during the semester 2018A in the frame of the program {\tt CAT18A-12}. All spectroscopic observations were obtained using the multi-object spectrographs Device Optimized for the LOw RESolution (DOLORES) at the Telescopio Nazionale Galileo (TNG) and Optical System for Imaging and low-Intermediate-Resolution Integrated Spectroscopy (OSIRIS) at the Gran Telescopio Canarias (GTC), both located at the Roque de los Muchachos Observatory (ORM) in La Palma (Spain). Details about the imaging and spectroscopic procedures can be found in \cite{paper4} and \cite{paper5}.

In total, 94 sources were observed spectroscopically, 55 at the GTC and 39 at the TNG. We obtained good quality data to estimate the velocity dispersion for 82 clusters, which corresponds to a success rate of $87$\,\%. The mean (median) redshift of this data set is $z_{spec} = 0.41\ (0.39)$ and the mean (median) number of galaxy members for these clusters is $ N = 26\ (22)$. All the spectroscopic results of these observations are presented here for the first time. Individual measurements for all cluster members (approximately 1400 redshifts in total) will be published online, and included in the VO.

\subsection{ITP13 data set}
\label{sec:itp13}

In addition, we use part of the {\tt ITP13} sample described in \cite{SRpsz1}. This sample consists of 61 PSZ1 clusters, from which 47 of them are also included in the PSZ2 catalog. The observations of these objects were performed during four semesters in the framework of the International Time Program {\tt ITP13B-08}, a similar program to the {\tt LP15} but for the PSZ1 catalog. We include these 47 objects in our analysis finding a mean (median) redshift of $z_{spec} = 0.37\ (0.31)$ and  a mean (median) number of galaxies members of the clusters $ N = 19\ (17)$. Out of those 47 objects, 43 of them are contained inside the PSZ2-North sample.
We complement the individual cluster member catalogs from the two data sets described above using spectroscopic data from the Sloan Digital Sky Survey \citep[SDSS,][]{SDSS} Data Release (DR) 14, when available.

It is important to clarify here that there might be slight differences between the velocity dispersion estimates quoted in this paper and those from \cite{SRpsz1}. Although the individual velocity catalogs used to estimate the velocity dispersion are the same, the methodology is not exactly identical, as described in detail in Section~\ref{sec:vd}.

\subsection{SDSS data}
\label{sec:sdss}

SDSS archival data give us a unique opportunity to enlarge our original sample. We retrieve every spectroscopic redshift within 15$\arcmin$ from the \planck\ nominal pointing for all the PSZ2 objects inside the SDSS footprint. For the cases with $z_{\rm spec}<0.1$, we expand this region to 30$\arcmin$ radius to obtain as much cluster members as possible.

We identify 259 galaxy clusters following this procedure. In nine cases, the object found does not fulfill the criteria to be considered an optical counterpart of the corresponding SZ source.
Those criteria are explained in detail in \cite{paper4} and \cite{paper5}. The mean (median) redshift of this data set is $z_{spec} = 0.22\ (0.19)$, and the mean (median) number of galaxies members of the clusters is $ N = 43\ (21)$.

\subsection{Other PSZ2 clusters}
\label{sec:others_psz2}

Table~\ref{tab:vd} also includes six GCs that do not belong to the PSZ2-North sample, but they are inside PSZ2. These objects are PSZ2 G021.02$-$29.04, PSZ2 G027.77$-$49.72-A, PSZ2 G027.77$-$49.72-B, PSZ2 G171.08$-$80.38, PSZ2 G208.57$-$44.31 and PSZ2 G270.78$+$36.83. They were observed for a different project but inside the {\tt LP15} program, so for this reason they are described here. As they do not form part of the PSZ2-North sample, they will not be considered for the characterization of the scaling relation. These objects are listed as "Others in PSZ2" (column 3) in Table~\ref{tab:sample}. 

\subsection{Beyond the PSZ2 sample}
\label{sec:bey_psz2N}

As anticipated above, during this program we have characterized 26 new clusters or groups that can not be formally associated with the PSZ2 detection because they do not fulfill the matching criteria for being considered the optical counterpart. These objects are presented in Table~\ref{tab:vd_others}, including their velocity dispersion, dynamical mass, number of members and redshift. As they are not associated with any SZ source, they cannot be used for the characterization of the scaling relation in section
~\ref{sec:sr}. They are listed as "Beyond PSZ2" in Table~\ref{tab:sample}.


\section{Velocity dispersion estimates}
\label{sec:vd}

Here we present the methodology and results for the estimation of the velocity dispersion for those 362 objects confirmed as the optical counterparts of SZ sources in the PSZ2 catalog (columns two and three in Table~\ref{tab:sample}). Table~\ref{tab:vd} shows the results for these GCs, and is organized as follows. Columns 1 and 2 are the official ID number and the \planck\ name in the PSZ2 catalog. Columns 4 and 5 are the J2000 coordinates of the BCG when present; otherwise, the geometrical center of the GC is provided. Columns 5 and 6 give the number of spectroscopic members retrieved. Columns 7, 8 and 9 provide the mean spectroscopic redshift of the cluster and, when available, the BCG's. Columns 10 and 11 are our velocity dispersion estimates. Columns 12, 13 and 14 present the dynamical and SZ mass estimates. Column 15 indicates whether the object was used in Sect.~\ref{sec:mass}. Column 16 lists the data set from where the cluster was extracted (see column 1 in Table~\ref{tab:sample}).

We also publish 26 clusters and groups found while studying the PSZ2 catalog, that are not associated with any SZ source, due to either their large distance from the \planck\ pointing, or to their low mass. They are presented in Table~\ref{tab:vd_others}, which is structured in a similar way as Table~\ref{tab:vd}. The difference is that these clusters and groups are not associated with any SZ source, so instead of naming them with the \planck\ name, we simply quote the field around which they were found. 

We follow the procedure outlined in \cite{Biases} to estimate the velocity dispersion. The authors demonstrate (using hydro-dynamical simulations) that the estimation of the velocity dispersion is biased in the low number of galaxies regime. They present a functional form, depending on the number of galaxies used, to correct for this effect \citep[see Eq. 11,][]{Biases}. They also show that the aperture sub-sampling is a source of error, and provide a recipe to correct for this effect. Finally, the authors note that the appropriate value of the clipping in the line-of-sight velocity field to minimize the presence of interlopers is 2.7. We adopt this value in our analysis. 

\begin{figure}
\centering \includegraphics[width=\columnwidth]{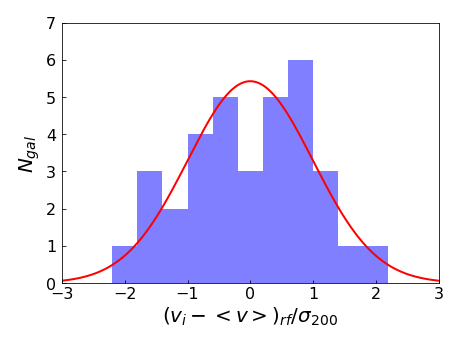}
\caption{Example of the distribution of galaxies in PSZ2 G009.04$+$31.09 as a function of the rest frame difference in radial velocity to the mean radial velocity of the cluster. In blue are represented the cluster members used to estimate the velocity dispersion. The red line represents the normal distribution expected for the estimated velocity dispersion of $\sigma_{200} = 1068$\,km\,s$^{-1}$. }
\label{fig:velhist}
\end{figure}

\begin{figure}
\centering \includegraphics[width=\columnwidth]{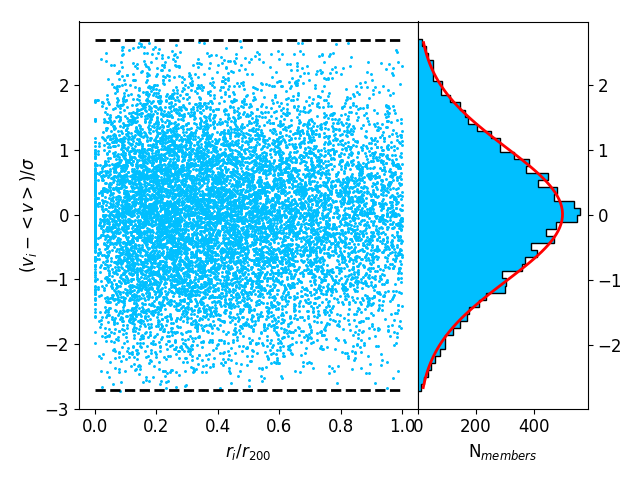}
\caption{Projected phase space and velocities histogram distribution for all the 11867 galaxy members in our sample. Member velocities are normalized to the mean cluster velocity dispersion, whereas the distance to the center of the cluster is normalized to the value of $r_{200}$ in each cluster. Horizontal black dashed lines are the 2.7$\sigma$ clip. The red line represents a Gaussian (normal) fit to the velocity histogram with $\sigma=1$. } 
\label{fig:caustic}
\end{figure}

We obtain the velocity dispersion in two steps. We make a first estimate using an iterative $\sigma-$clipping method and then we apply the corrections to the estimator. For the iterative $\sigma-$clipping method, we use a clip of 2.7$\sigma$ and a cut in aperture of 1 or 1.5 $\times r_{200}$ which is included inside the clipping. Once we have obtained this first estimate, we apply the corrections due to the used estimator and the aperture. In this paper, we choose the gapper estimator \citep{gapper}, as it is the one with the least dependence on the number of galaxies \citep{Biases}. Figure~\ref{fig:velhist} shows an example of the final velocity histogram of the cluster members for a particular case in our sample. Figure~\ref{fig:caustic} shows the stacked distribution of all the galaxies in the phase space, for all clusters.

As mentioned in Sect.~\ref{sec:sample} and above, we use two different apertures when selecting the cluster members. The main reason is to evaluate the possible bias introduced by the presence of interlopers inside the individual cluster catalogs. The interlopers are an important cause of uncertainty when estimating the velocity dispersion of a cluster as shown by many authors in the literature \citep[see e.g. ][]{Mamon11, Saro13, Wojtak18, Pratt19}. For each cluster, we present both the values for the case of an aperture of $1.5\times r_{200}$ as well as for $r_{200}$. We note that when restricting the aperture limit to $r_{200}$, we find 36 GCs less, due to the drop in the number of members, as the minimum number of members that we consider to estimate the velocity dispersion is seven.

Among those 362 presented counterparts, five are what we call a "multiple detection". This means that there are more than one cluster associated with the SZ signal. In addition, there are 16 objects that are clearly sub-structured, so their velocity dispersion estimates should not be trusted, as they probably overestimate the true underlying velocity. We do not use these objects when characterizing the $M_{SZ}-M_{dyn}$ scaling relation. 

Unfortunately, not all of the clusters we found are associated with the SZ emission. There might be low mass systems or objects too distant from the SZ peak to be considered the counterpart. These clusters are not used for the calibration of the scaling relation. We also mark these objects in Table~\ref{tab:vd_others}.


\section{Mass estimates}
\label{sec:mass}

In the following sub-sections, we describe our methodology to obtain the dynamical and SZ masses. These masses are used in Sect.~\ref{sec:sr} to characterize their scaling relation and to obtain the bias parameter $(1-b)$ which is of enormous importance for cosmological studies.

The mass estimates are presented in Columns 12, 13 and 14 of Table~\ref{tab:vd}. Column 15 indicates whether an object is used in Sect.~\ref{sec:sr} to characterize the scaling relation. Columns 5, 6 in Table~\ref{tab:sample} show the total number of GCs in each data-set used for the estimation of the mass bias parameter. 

\subsection{$M_{dyn}$ estimates}
\label{sec:Mdyn}

Estimating the dynamical mass of a cluster is not a simple task. As shown in \cite{GCMRPI}, the estimation of the mass using a low number of cluster members is problematic. For this reason, we use the method described in \cite{Biases}, where the authors study the behavior of several velocity dispersion and dynamical mass estimators using hydro-dynamical simulations in the low number of galaxies regime. They demonstrate that the estimation of the velocity dispersion is biased in this regime, and propose a functional form that depends on the number of galaxies used to correct for this fact \citep[Eqs.\,15 and 16,][]{Biases}.

The scaling relation used for the estimation of the dynamical mass is Eq. \ref{eq:munari} from \cite{Munari13}:

\begin{equation} \label{eq:munari}
\frac{M_{200}^{dyn}}{10^{15}\textup{M}_\odot} = \left(\frac{\sigma_{200}}{A}\right)^\frac{1}{\alpha},
\end{equation}
where $A = 1177.0$\,km\,s$^{-1}$ and $\alpha = 0.364$. 
We note that these parameters were obtained for a velocity dispersion calculated using the biweight estimator \citep{Beers90}. Thus, for consistency, we convert our velocity dispersion estimates to that of the biweight, following the recipe in \cite{Biases}. After applying the corrections to $M_{200}^{dyn}$ due to the number of cluster members, we convert this mass into $M_{500}^{dyn}$, so we can compare it to $M_{500}^{SZ}$. This last step is performed using the python package {\it NFW}\footnote{https://github.com/joergdietrich/NFW} which implements the Navarro, Frenck and White \citep{NFW} halo profile using the concentration parameter from \cite{Duffy08}:
\begin{equation}\label{eq:duffy}
c_{200} = 5.71\cdot(1+z)^{-0.47}\left(\frac{M_{200}}{2 \times 10^{12}h^{-1}M_{\odot}}\right)^{-0.084}.
\end{equation}

The uncertainties on $M_{500}^{dyn}$ are based on the expected variance of our estimator as a function of the number of galaxies, as shown in Figure~8 in \cite{Biases}. Those curves can be fitted to an equation of the type:
\begin{equation}
\Delta M_{200}^{dyn}=M_{200}^{dyn}\sqrt{\frac{c}{4(N_{gal}-1)^\alpha}}.
\end{equation}
Finally, to obtain the uncertainity in $\Delta M_{500}^{dyn}$ from $\Delta M_{200}^{dyn}$, we use a quadratic propagation of the error, but the uncertainty in the concentration parameter is not propagated.

\subsection{$M_{SZ}$ estimates}
\label{sec:Msz}

The \planck\ collaboration provides, for every SZ source in the PSZ2 catalog, an array of masses as a function of redshift, $M_{500,nc}^{SZ}(z)$. These values were obtained by breaking the size-flux degeneracy of the \planck\ measurements using a prior relating the SZ flux ($Y_{500}$) and the cluster size ($\theta_{500}$). In turn, this cluster size is connected to the total mass for a given redshift $z$. For each cluster in our sample, we interpolate these arrays to our measured redshift, and extract their SZ mass. Further details about the procedure to obtain $M_{500,nc}^{SZ}(z)$ can be found in \cite{2014A&A...571A..20P} and \cite{PSZ2}.

These SZ masses suffer from Eddington bias \citep[see e.g.][]{vanderBurg16}, specially in the low signal-to-noise regime.
Fig.~5 in \cite{vanderBurg16} shows the magnitude of Eddington bias as a function of the signal-to-noise ratio in the PSZ2 catalog for different redshifts. They estimate this Eddington bias by simulating a list of masses and redshifts following the \cite{Tinker08} halo mass function and the redshift-dependent co-moving volume element for their assumed cosmology.  We use that figure to create a hyper-surface and apply a 3D interpolation technique in order to correct our SZ masses for this effect and obtain the final $M_{500}^{SZ}$. We note that this treatment is an approximation, as the correction for each cluster is purely statistical \citep[see e.g. Appendix A in][for an illustration of the effect of this type of statistical bias]{Mantz10}. For this reason, our individual corrected masses should be seen as an approximation. Nevertheless, the overall mass bias for the full sample should be correctly estimated.


\section{$M_{SZ}-M_{dyn}$ scaling relation}
\label{sec:sr}

In this section we present and discuss the scaling relation between SZ and dynamical masses for a statistically representative sample of the PSZ2 catalog. Starting from the complete list of clusters presented in Sect.~\ref{sec:sample}, we use two additional criteria to remove objects from the list. We exclude: (i) GCs that are clearly sub-structured, as the estimation of the dynamical mass is probably overestimated in this case, and (ii) those presenting multiple counterparts, as using \planck\ data alone and due to the beam size, it is not possible to disentangle the individual contribution of each cluster to the total SZ flux. 

After applying these two exclusion criteria, our final sample adopted for the computation of the scaling relation contains 297 PSZ2 clusters, all of them with members selected within $1.5 \times r_{200}$. Column 5 in Table~\ref{tab:sample} shows the distribution of those objects in the three data sets considered in Sect.~\ref{sec:sample}. Fig.~\ref{fig:bfull} presents the scaling relation obtained for those 297 objects. From here, our main goal is to find the so-called mass bias factor $(1-b)$, which accounts for any difference between the true mass and the SZ mass proxies ($Y_{500}^{SZ}$, $\theta_{500}^{SZ}$) used to establish the scaling relations. We define this bias as
\begin{equation}\label{eq:bsz}
M_{500}^{\rm SZ} = (1-b)\ M_{500}^{\rm true}.
\end{equation}

As explained in Sect.~\ref{sec:vd}, we are not able to correct for all the physical effects potentially causing a bias when estimating the true velocity dispersion of the clusters. This leads to a bias between the true mass and the dynamical mass estimates. The main source of error are possibly the interlopers inside our sample. For this reason, we define the dynamical mass bias factor $(1 - b_{dyn})$ as
\begin{equation} \label{eq:bdyn}
M_{500}^{\rm dyn} = (1-b_{\rm dyn})\ M_{500}^{\rm true}.
\end{equation}
Combining the equations \ref{eq:bsz} and \ref{eq:bdyn}, we obtain
\begin{equation} \label{eq:B}
M_{500}^{\rm SZ} = (1-B)\ M_{500}^{\rm dyn},
\end{equation}
where
\begin{equation} \label{eq:Bfrac}
(1-B) \equiv \frac{(1-b)}{(1-b_{\rm dyn})}.
\end{equation}

We study this last bias $(1-B)$ in our scaling relation. In principle, we would expect that this bias $(1-B)$ represents a lower bound to $(1-b)$, the reason being that the presence of interlopers generally produces an overestimation of the velocity dispersion, and thus of the dynamical mass \citep[e.g.][]{Biases}. However, and to estimate the real impact of interlopers in our sample, we repeat the same analysis with a smaller sub-sample of 261 clusters obtained by reducing the aperture when selecting the cluster members to $r_{200}$. As shown below, we find consistent results in this case, suggesting that the impact of interlopers for our particular sample is minimal (or at least smaller than our statistical error), as anticipated in \cite{Biases}. Column 6 in Table~\ref{tab:sample} shows the distribution of objects through the data sets of this smaller sub-sample.

\subsection{Regression method}
\label{sec:regmethods}

Here, we characterize with realistic simulations, matching the statistical properties of our sample, various regression methods. These simulations follow the very same procedure as the real data, using the same number of galaxies for each cluster to estimate the velocity dispersion and the dynamical mass uncertainties. These simulations are detailed in Appendix~\ref{app:B}. We explore two possibilities. First, we consider the simplest case of fitting for a global bias. However, as there are hints that suggest a possible mass dependence of the mass bias, we also explore a fit to a power law in mass to account for this dependence, using as pivot scale $6\times10^{14}M_{\odot}$, to be able to consistently compare our results with other works in the literature \citep[e.g. ][]{2016A&A...594A..24P}. The parametric form of the fitting function in this second case is given by:
\begin{equation} \label{eq:Bnorm}
\frac{M_{500}^{SZ}}{6\times10^{14}M_{\odot}} = (1-B)\ \left( \frac{M_{500}^{dyn}}{6\times10^{14}M_{\odot}} \right) ^{\alpha}.
\end{equation}
To be clear, when fitting this power law, the result of the mass bias is estimated at this given mass of $6\times10^{14}M_{\odot}$.

Due to the large uncertainties in the dynamical mass estimates for our sample, all the methods that we have tested do not behave well and give completely biased outputs. For this reason, we do the linear regression in the logarithmic space, fitting for the relation
\begin{equation} \label{eq:lnBnorm}
\ln{\left( \frac{M_{500}^{SZ}}{6\times10^{14}M_{\odot}}\right) = {\alpha}}\  \ln{\left( \frac{M_{500}^{dyn}}{6\times10^{14}M_{\odot}} \right)} + ln{\left(1-B\right)}, 
\end{equation}
where $\alpha$ and $ln\left(1-B\right)$ are the slope and the intercept, respectively.
It is important to note that in our limit of large dynamical mass errors, these are also considerably greater than the expected intrinsic scatter of the relation $\sigma_{ln\ M}=0.096$ \citep{2014A&A...571A..20P}.

We study the dependence of the mass bias with mass by fitting the slope in Equation~\ref{eq:lnBnorm}. To do so, we tested five different regression methods which are usually applied in the literature: Orthogonal Distance Regression (ODR), Nukers \citep{Tremaine02},  Maximum Likelihood Estimator with Uniform prior (MLEU), Bivariate Correlated Errors and intrinsic Scatter \citep[BCES,][]{BCES}, and the Complete Maximum Likelihood Estimator \citep[CMLE,][]{Kelly07}. 
All of them are described in detail in Appendix~\ref{app:B}, and applied to simulated data based on the noise distributions and range of masses in our real sample.
Our results show that all methods are biased, although some of them are more robust and less affected by errors.  In general, those methods that take into account the intrinsic scatter in an explicit way (BCES, MLEU and CMLE) fail in the recovery of this parameter, probably due to the fact that the error measurements are of the order of or larger than the intrinsic scatter. For our sample, both the ODR and the Nukers method appear to be more robust, showing a small bias of  approximately $7\%$ when recovering the intercept, being the slope well recovered. Finally, we also tested the case of no mass dependence in the scaling relation, i.e. fixing the slope to one. We tested three methods in this limit: ODR, Nukers and MLEU. The results are similar to those from the complete regression. The ODR and Nukers methods are biased by approximately $7$\,\% in the intercept, while the MLEU fails to recover both input values. This bias in ODR and Nukers of 7\,\% in the intercept is robust, independently of the input value adopted for the mass bias in the simulations (we tested values from 0.6 to 1.2). 

For completeness, we also studied other cases to understand the range of applicability of the different methods, always keeping the same number of clusters as in our sample. In particular, we considered the case of decreasing significantly the errors in the dynamical mass, and also a case with no intrinsic scatter in the simulated data.
If the statistical errors in the dynamical mass measurements are significantly decreased (by a factor of 10), then all methods are able to recover both parameters, regardless whether there is an intrinsic scatter in the simulation.  
These tests strongly suggest that the main source of bias of the methods are the large errors of our data. 

The two methods that better recover the parameters in our case are Nukers and ODR, when fixing and without fixing the slope. However, as shown in Appendix~\ref{app:B}, both methods present a small bias in the intercept that has to be corrected. In this paper, we use Nukers  as the reference method. It gives practically the same results as the ODR, while it also gives an estimation of the intrinsic scatter. 
Our simulations also show that this method provides consistent results for both cases of fixing the slope or not. 
When quoting our final values, we correct our Nukers results accounting for the bias in the intercept ($7$\,\%), and add a systematic uncertainty due to this bias.

\subsection{Results}
\label{sec:bias}        

\begin{table*}
\centering
\caption{Results for the mass bias using both sub-samples 1 and $1.5 \times r_{200}$. We present the direct results of the Nukers method for both the case of fixed slope ($\alpha=1$) and free-slope, and also the corresponding bias-corrected values. See text for details. }
\begin{tabular}{c c c c c}
\hline \hline
\noalign{\smallskip}
\multirow{2}{*}{Method}    & \multicolumn{2}{c}{$1.5\times r_{200}$} & \multicolumn{2}{c}{$1\times r_{200}$}     \\
\cline{2-3}
\cline{4-5}
\noalign{\smallskip}
 & $(1-B)$ & $\alpha$ & $(1-B)$ & $\alpha$ \cr

\hline
\noalign{\smallskip}                                                                 
 \multirow{2}{*}{Nukers} & 0.850 $\pm$ 0.040  &  1.000             & 0.841 $\pm$ 0.040  &  1.000             \\
                         & 0.889 $\pm$ 0.065  &  1.167 $\pm$ 0.125 & 0.875 $\pm$ 0.067  &  1.145 $\pm$ 0.121  \\

\hline
\noalign{\smallskip}                                                                 
Nukers                   & 0.80 $\pm$ 0.04 $\pm$ 0.05 &  1.00             & 0.79 $\pm$ 0.04 $\pm$ 0.05  &  1.00                       \\
corrected                & 0.84 $\pm$ 0.07 $\pm$ 0.05 &  1.17 $\pm$ 0.13  & 0.83 $\pm$ 0.07 $\pm$ 0.05  &  1.15 $\pm$ 0.13 \\

\hline
\end{tabular}
\label{tab:bias}

\end{table*}

Table~\ref{tab:bias} shows the results for the fitting of Eq.~\ref{eq:lnBnorm} using the Nukers method described above, and for both samples. The statistical errors quoted in this table are computed with a bootstrapping technique. 

The value of the slope is (in both samples) a bit more than 1-$\sigma$ away from one, which might be indicative of a possible dependence of the $(1-B)$ with the mass, but  the results are not significant enough to make that claim. 
For comparison, \cite{vonderLinden14} and \cite{Hoekstra15} find a slope around 0.7, which goes in the direction of inverse dependence of the mass bias with mass. Their results were obtained using the CMLE method \citep{Kelly07}. 
We note that in our particular case, the simulations show that this method presents a significant bias, of approximately $20$\,\% (see appendix~\ref{app:B}). A direct evaluation of the slope using CMLE gives $\alpha=0.70\pm0.06$ for our sample, but after the bias correction, this number moves up to $\alpha=0.88\pm 0.07$, which is less than 2-$\sigma$ from unity. 

\begin{figure}
\centering \includegraphics[width=\columnwidth]{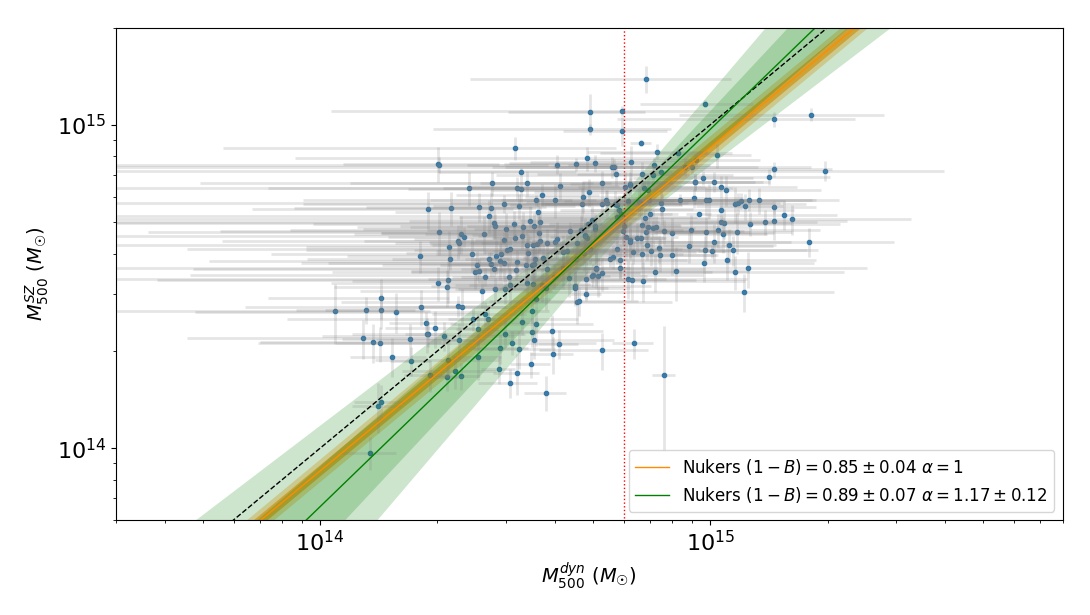}
\caption{Scaling relation for the sample of 297 PSZ2 clusters ($1.5\times r_{200}$ sample). The dashed black line shows the $1:1$ line. The orange line represents our best fit using the Nukers method with $\alpha=1$ (see text for details). The green line is the fit using the complete Nukers method, for a free slope. The shaded regions represent the 1- and 2-$\sigma$ errors of the reconstructed parameters. The vertical red dotted line corresponds to the pivot  mass of $6\times 10^{14} M_{\odot}$. }
\label{fig:bfull}
\end{figure}

We find no significant difference in the $(1-B)$ mass bias when considering different samples, suggesting that the effect of the interlopers in the region $1-1.5\ \times r_{200}$ is smaller than the quoted statistical error, as expected. For this reason, we restrict the following analysis to the case of the full sample ($1.5\ \times r_{200}$). The results for the mass bias using this sample are shown in Fig.~\ref{fig:bfull}.

\begin{table*}
\centering   
\tiny
\caption{Results for the mass bias using the sample ($1.5 \times r_{200}$), when restricting the analysis to those clusters with total number of spectroscopic members $N$ in a certain range or interval. See text for details.  }
\begin{tabular}{c c c c c c c c c}
\hline \hline
\noalign{\smallskip}

                             & \multicolumn{2}{c}{$N>15$}                 & \multicolumn{2}{c}{$N>20$}                 & \multicolumn{2}{c}{$N>30$}                   & \multicolumn{2}{c}{$N>50$} \\

\cline{2-3}
\cline{4-5}
\cline{6-7}
\cline{8-9}
\noalign{\smallskip}

                             & $(1-B)$          & $\alpha$                & $(1-B)$          & $\alpha$               & $(1-B)$            & $\alpha$                 & $(1-B)$         & $\alpha$              \cr

\hline                                                                                                                                    
\noalign{\smallskip}                                                                                                                      
 \multirow{2}{*}{Nukers}      & 0.845 $\pm$ 0.041 &  1.000                & 0.841 $\pm$ 0.043   &  1.000              & 0.824 $\pm$ 0.045    &  1.000                 & 0.815 $\pm$ 0.051 &  1.000                 \\
                              & 0.881 $\pm$ 0.067 &  1.154 $\pm$ 0.127    & 0.880 $\pm$ 0.070   &  1.163 $\pm$ 0.131  & 0.859 $\pm$ 0.080    &  1.143 $\pm$ 0.155     & 0.860 $\pm$ 0.100 &  1.175 $\pm$ 0.181       \\
\hline
\noalign{\smallskip}
$N_{clusters}$               & \multicolumn{2}{c}{214}                    & \multicolumn{2}{c}{164}                    & \multicolumn{2}{c}{100}                      & \multicolumn{2}{c}{55} \\

\hline
\end{tabular}
\label{tab:bias_Ncut1.5}

\end{table*}

We investigate the robustness of the results when selecting only those clusters with smaller error bars in the determination of the dynamical mass. For this, we have repeated our analysis using different selections according to the number of cluster members used to determine the velocity dispersion (parameter $N_{1.5}$ in Table~\ref{tab:vd}). 
Table~\ref{tab:bias_Ncut1.5} shows the results when restricting our sample to those clusters with $N_{1.5} > 15,\ 20,\ 30$, and $50$. The values of the mass bias and the slope are consistent with each other in all cases. However, we note that there is a marginal trend (smaller than 1-$\sigma$) towards lower values of the bias for the sub-samples with more cluster members. We can understand this trend by noting that those clusters with more members are, on average, less massive as they are mostly low-redshift systems. As there is a marginal detection of a slope $\alpha > 1$, then we would expect low-mass clusters to present a lower $(1-B)$.

\subsubsection{Eddington bias}
\begin{table}
\centering
\caption{Nukers $(1-B)$ estimates before and after the Eddington bias correction in signal-to-noise ratio bins for the sample with aperture cut $1.5 \times r_{200}$.}
\begin{tabular}{c c c}
\hline \hline
\noalign{\smallskip}

\multirow{2}{*}{SNR bin}    &  \multicolumn{2}{c}{$(1-B)$} \\
\cline{2-3}
\noalign{\smallskip}
                        &      Before       &     After   \\
\hline
\noalign{\smallskip}

 \hspace{1cm}SNR    <   4.97  & 1.009 $\pm$ 0.075 & 0.900 $\pm$ 0.071     \\
 4.97 $\leq$ SNR    <   5.57  & 0.919 $\pm$ 0.065 & 0.818 $\pm$ 0.056     \\
 5.57 $\leq$ SNR    <   6.35  & 0.789 $\pm$ 0.058 & 0.730 $\pm$ 0.053     \\
 6.35 $\leq$ SNR    <   8.26  & 0.912 $\pm$ 0.062 & 0.873 $\pm$ 0.057     \\
 \hspace{1cm}SNR $\geq$ 8.26  & 0.882 $\pm$ 0.082 & 0.876 $\pm$ 0.079     \\

\hline
\end{tabular}
\label{tab:SNRbins}
\end{table}

The effect of the Eddington bias correction is shown in Table~\ref{tab:SNRbins}. Clusters are distributed in five bins, keeping the same number of objects per bin. We restrict the analysis here to the case of a fixed slope ($\alpha=1$) as the size of the errors likely prevents to derive any constrain on the slope if left free. As expected, the correction applied reduces the mass bias between $10\%$ and less than $1\%$, depending on the SNR of the \Planck\ SZ detection. We note that the central bin (5.57 < SNR $\leq$ 6.35) does not follow the trend of the others, but it is still less than 2-$\sigma$ away from the mean value for the full sample.

\subsubsection{Redshift dependence}

\begin{table}
\centering
\caption{Nukers $(1-B)$ estimates for different bins in redshift for the sample with aperture cut $1.5 \times r_{200}$.}
\begin{tabular}{c c c c}
\hline \hline
\noalign{\smallskip}

Redshift bin    &  $(1-B)$ \\

\hline
\noalign{\smallskip}

\hspace{1.2cm}z     <    0.107 & 0.810 $\pm$ 0.059  \\
 0.107 $\leq$ z     <    0.200 & 1.013 $\pm$ 0.061  \\
 0.200 $\leq$ z     <    0.292 & 0.784 $\pm$ 0.052  \\
 0.292 $\leq$ z     <    0.379 & 0.799 $\pm$ 0.055  \\
\hspace{1.2cm}z  $\geq$  0.379 & 1.038 $\pm$ 0.068  \\

\hline
\end{tabular}
\label{tab:zbins}
\end{table}

Table~\ref{tab:zbins} presents the results for the mass bias in five different redshift bins with the same number of objects. As in the previous study, there are not enough clusters in each bin to make the regression with a free varying slope, so we restrict the analysis again to the case of fixed ($\alpha=1$) slope. We find that three of the five bins show consistent results with the mean bias for the full sample. However, there are two bins ($0.107 \leq z  < 0.200$ and $z \geq 0.379$), which are inconsistent with the mean bias at the level of approximately 2.7-$\sigma$. We refer to them as second and fifth bins. 

We have carried out several tests to explain the origin of these two outliers, but none of the analyses are conclusive. First, we have explored if this difference could be ascribed to a significantly different mean mass in the bin. In principle, the low redshift bins could span a larger range of masses due to the survey selection function \citep[see Fig.~26 in][]{PSZ2}. The median mass values that we find for those five bins are $3.6$, $4.3$, $7.1$, $6.7$ and $5.9 \times 10^{14}\text{M}_{\odot}$, respectively. These values do not show any specific trend that could explain the two outliers.   
This is also the case for the mean number of galaxies (90, 27, 19, 18 and 22) and the mean SNR (8.8, 7.5, 7.7, 6.1 and 5.9) in each of the five bins. 
We have also divided the two anomalous bins in two new sub-bins in redshift, SNR and number of galaxies. There is no appreciable difference in any sub-bin for the case of $0.107 \leq z  < 0.200$. However, when we do the same for the fifth bin ($z \geq 0.379$), it seems that the outlier here might be due to the high redshift clusters. This is in agreement with the result showed before of a small dependence of the mass bin with mass. 

Thus, we cannot find a simple explanation for the outlier in the second redshift bin ($0.107 \leq z  < 0.200$), and if confirmed with better statistics, this could be ascribed to a real physical effect.
For comparison, we note that \cite{SRpsz1} also find a similar outlier in the same redshift bin when using PSZ1 clusters only.

\section{Summary and conclusions}
\label{sec:con}

This is the third and last paper in a series describing the observational program {\tt LP15}. Here, we presented the spectroscopic data of the full program. In total, 94 PSZ2 sources were observed, 55 at the GTC and 39 at the TNG. We were able to estimate the velocity dispersion for 82 clusters. In addition, we used 47 clusters from the {\tt ITP} sample and 259 clusters from the SDSS archival data to build a statistically representative sample of the PSZ2 in the northern hemisphere (PSZ2-North).

We presented the velocity dispersion and dynamical mass of 362 objects confirmed as optical counterpart of a PSZ2 source, 356 from the PSZ2-North sample and nine from outside. We also discuss 26 clusters and groups that do not fulfill the matching criteria to be a counterpart of the SZ signal.

The combination of {\tt LP15}, {\tt ITP} and SDSS samples yields a total sample of 297 galaxy clusters that can be used for the characterization of the scaling relation $M_{SZ}-M_{dyn}$ for the PSZ2 catalog. This sample represents the largest set of SZ selected clusters for which SZ and dynamical masses are available. It is, in fact, the largest sample used to determine the mass bias using dynamical mass estimates.

Based on a set of realistic simulations which are representative of the actual noise level in our sample, we have selected the Nukers method as the least-biased regression method to extract the scaling relation. 
After correcting for the statistical bias of the regression method and the Eddington bias of the sample, we find the mass bias to be $(1-B) = 0.80 \pm 0.04$ (stat)$\pm 0.05$ (sys). Assuming $(1-b_{dyn})=1$, our value for $(1-b)$ is in agreement with previous studies \citep{Ruel14,Hoekstra15,Battaglia16,Sereno17,PennaLima17,Medezinski18,Miyatake19,SRpsz1}, and do not solve the tension in the cosmological parameters ($\Omega_{\rm m}$--$\sigma_8$ plane) between the CMB measurements and the cluster count analyses \citep{2016A&A...594A..24P,2018A&A...614A..13S,2018arXiv180706209P,2019MNRAS.483.3459R}, which requires lower values for $(1-b)$.

We note that \citet{SRpsz1} present a similar analysis to the one carried out in this paper, but for 207 PSZ1 clusters. Their value is $(1-B)=0.83\pm0.07\pm0.02$, which is in complete agreement with our result. In that paper, there is a detailed comparison with similar works in the literature, including the one presented here.

Finally, we only find marginal evidence of a possible dependence of the mass bias with mass. Our fitted slope is $\alpha = 1.17\pm0.13$ which is 1.3-$\sigma$ away from the mass-invariant relation.

\begin{acknowledgements}

This article is based on observations made with the Gran Telescopio Canarias operated by the Instituto de Astrofisica de Canarias, the Isaac Newton Telescope, and the William Herschel Telescope operated by the Isaac Newton Group of Telescopes, and the Italian Telescopio Nazionale Galileo operated by the Fundacion Galileo Galilei of the INAF (Istituto Nazionale di Astrofisica). All these facilities are located at the Spanish Roque de los Muchachos Observatory of the Instituto de Astrofisica de Canarias on the island of La Palma. This research has been carried out with telescope time awarded for the program 128-MULTIPLE-16/15B. During our analysis, we also used the following databases: the SZ-Cluster Database operated by the Integrated Data and Operation Center (IDOC) at the IAS under contract with CNES and CNRS and the Sloan Digital Sky Survey (SDSS) DR14 database. Funding for the SDSS has been provided by the Alfred P. Sloan Foundation, the Participating Institutions, the National Aeronautics and Space Administration, the National Science Foundation, the U.S. Department of Energy, the Japanese Monbukagakusho, and the Max Planck Society. This work has been partially funded by the Spanish Ministry of Science and Innovation under the projects ESP2013-48362-C2-1-P, AYA2014-60438-P, AYA2017-84185-P and PID2020-120514GB-I00. AS and RB acknowledge financial support from the Spanish Ministry of Science and Innovation under the Severo Ochoa Programs SEV-2011-0187 and SEV-2015-0548.

\end{acknowledgements}

\bibliographystyle{aa} 
\bibliography{draft.bib}

\appendix
\section{Tables with the main results of the paper}
\label{app:A}

\begin{landscape}
\begin{table}
\caption{362 PSZ2 optical counterparts presented in this paper.}
\centering
\tiny
\begin{tabular}{c c c c c c c c c c c c c c c c c}
\hline\hline                        
\noalign{\smallskip}
 ID$^1$ & NAME & R.A. & Dec. & N$_{1.5}$ & N$_1$ & z$_{1.5}$ & z$_1$ & z$_{BCG}$ & $\sigma_{200,1.5}$ & $\sigma_{200,1}$ & $M^{dyn}_{500,1.5}$ & $M^{dyn}_{500,1}$ & $M^{SZ}_{500}$ & Flag$^2$ & Data set \\
\noalign{\smallskip}
 &      &      &      &           &       &           &       &           & (km\,s$^{-1}$)     & (km\,s$^{-1}$)   & $(10^{15}\textup{M}_\odot)$  &  $(10^{15}\textup{M}_\odot)$  & $(10^{15}\textup{M}_\odot)$ & & \\
\hline 
\noalign{\smallskip}
    1       & PSZ2 G000.04$+$45.13 & 15:16:58.08 & -01:06:38.8 &  38 &  15 & 0.118 & 0.118 & 0.1169 &  652 $\pm$ 74  &  488 $\pm$ 83  & 0.20 $\pm$ 0.05 & 0.10 $\pm$ 0.05 & 0.38 $\pm$ 0.04  &  0 & SDSS\\
    2       & PSZ2 G000.13$+$78.04 & 13:34:08.68 & +20:14:53.0 &  18 &  16 & 0.172 & 0.172 & 0.1730 & 1019 $\pm$ 158 & 1019 $\pm$ 166 & 0.68 $\pm$ 0.32 & 0.69 $\pm$ 0.35 & 0.51 $\pm$ 0.03  &  1 & SDSS\\
    7       & PSZ2 G002.42$+$69.64 & 14:04:08.04 & +15:40:09.7 &  11 &  11 & 0.183 & 0.183 & 0.1856 & 1243 $\pm$ 249 & 1248 $\pm$ 250 & 1.22 $\pm$ 0.85 & 1.24 $\pm$ 0.86 & 0.30 $\pm$ 0.04  &  1 & SDSS\\
   14$^{a}$ & PSZ2 G004.04$+$42.23 & 15:33:03.55 & -00:46:16.5 &  24 &  24 & 0.153 & 0.153 & 0.1518 & 1414 $\pm$ 187 & 1423 $\pm$ 188 & 1.63 $\pm$ 0.61 & 1.66 $\pm$ 0.62 & 0.31 $\pm$ 0.04  &  0 & SDSS\\
   21       & PSZ2 G006.49$+$50.56 & 15:10:54.11 & +05:46:54.8 & 223 & 173 & 0.078 & 0.078 &    -   & 1058 $\pm$ 61  & 1109 $\pm$ 54  & 0.75 $\pm$ 0.05 & 0.85 $\pm$ 0.07 & 0.71 $\pm$ 0.02  &  1 & SDSS\\
   25       & PSZ2 G006.84$+$50.69 & 15:10:59.21 & +05:58:57.9 & 226 & 166 & 0.078 & 0.078 &    -   & 1064 $\pm$ 64  & 1056 $\pm$ 52  & 0.76 $\pm$ 0.05 & 0.75 $\pm$ 0.06 & 0.17 $\pm$ 0.07  &  2 & SDSS\\
\hline 
\end{tabular}
\label{tab:vd}
\end{table}
\addtocounter{table}{-1}

\end{landscape}

\begin{landscape}
\begin{table}
\caption{Continued.}
\label{tab:vd2}
\centering
\tiny
\begin{tabular}{c c c c c c c c c c c c c c c c c}
\hline\hline                        
\noalign{\smallskip}
 ID$^1$ & NAME & R.A. & Dec. & N$_{1.5}$ & N$_1$ & z$_{1.5}$ & z$_1$ & z$_{BCG}$ & $\sigma_{200,1.5}$ & $\sigma_{200,1}$ & $M^{dyn}_{500,1.5}$ & $M^{dyn}_{500,1}$ & $M^{SZ}_{500}$ & Flag$^2$ & Data set \\
\noalign{\smallskip}
 &      &      &      &           &       &           &       &           & (km\,s$^{-1}$)     & (km\,s$^{-1}$)   & $(10^{15}\textup{M}_\odot)$  &  $(10^{15}\textup{M}_\odot)$  & $(10^{15}\textup{M}_\odot)$ & & \\
\hline 
\noalign{\smallskip} 
  204       & PSZ2 G050.40$+$31.17 & 17:20:10.83 & $+$27:42:14.8 &  19 &  16 & 0.160 & 0.160 & 0.1589 &  991 $\pm$ 154 & 1012 $\pm$ 165 & 0.63 $\pm$ 0.28 & 0.68 $\pm$ 0.35 & 0.40 $\pm$ 0.03  &  1 & SDSS\\
  210       & PSZ2 G052.35$-$31.98 & 21:18:42.70 & $+$00:32:06.1 &  11 &  -  & 0.275 &   -   & 0.2709 &  699 $\pm$ 144 &      -         & 0.24 $\pm$ 0.17 &      -          & 0.40 $\pm$ 0.05  &  2 & SDSS\\
  212       & PSZ2 G053.44$-$36.25 & 21:35:19.80 & $-$01:02:06.6 &  15 &  11 & 0.327 & 0.326 &    -   & 1146 $\pm$ 201 & 1112 $\pm$ 222 & 0.86 $\pm$ 0.46 & 0.83 $\pm$ 0.57 & 0.74 $\pm$ 0.05  &  1 & SDSS\\
  213       & PSZ2 G053.53$+$59.52 & 15:10:07.40 & $+$33:29:25.4 &  64 &  64 & 0.114 & 0.114 &    -   & 1106 $\pm$ 90  & 1106 $\pm$ 90  & 0.84 $\pm$ 0.15 & 0.84 $\pm$ 0.15 & 0.59 $\pm$ 0.02  &  1 & SDSS\\
  222       & PSZ2 G054.99$+$53.41 & 15:39:31.06 & $+$34:28:15.8 &  11 &   7 & 0.227 & 0.228 &    -   &  741 $\pm$ 152 &  784 $\pm$ 202 & 0.30 $\pm$ 0.21 & 0.38 $\pm$ 0.39 & 0.57 $\pm$ 0.04  &  1 & SDSS\\
  223       & PSZ2 G055.59$+$31.85 & 17:22:27.18 & $+$32:07:57.3 &  18 &  12 & 0.225 & 0.227 & 0.2233 & 1152 $\pm$ 185 & 1094 $\pm$ 209 & 0.92 $\pm$ 0.43 & 0.83 $\pm$ 0.54 & 0.78 $\pm$ 0.03  &  1 & SDSS\\
\hline 
\end{tabular}
\end{table}
\addtocounter{table}{-1}

\end{landscape}

\begin{landscape}
\begin{table}
\caption{Continued.}
\label{tab:vd3}
\centering
\tiny
\begin{tabular}{c c c c c c c c c c c c c c c c c}
\hline\hline                        
\noalign{\smallskip}
 ID$^1$ & NAME & R.A. & Dec. & N$_{1.5}$ & N$_1$ & z$_{1.5}$ & z$_1$ & z$_{BCG}$ & $\sigma_{200,1.5}$ & $\sigma_{200,1}$ & $M^{dyn}_{500,1.5}$ & $M^{dyn}_{500,1}$ & $M^{SZ}_{500}$ & Flag$^2$ & Data set \\
\noalign{\smallskip}
 &      &      &      &           &       &           &       &           & (km\,s$^{-1}$)     & (km\,s$^{-1}$)   & $(10^{15}\textup{M}_\odot)$  &  $(10^{15}\textup{M}_\odot)$  & $(10^{15}\textup{M}_\odot)$ & & \\
\hline 
\noalign{\smallskip}
  352       & PSZ2 G081.72$+$70.15 & 14:00:26.01 & $+$41:01:35.3 &  12 &  10 & 0.250 & 0.250 & 0.2497 &  749 $\pm$ 147 &  638 $\pm$ 134 & 0.30 $\pm$ 0.19 & 0.20 $\pm$ 0.15 & 0.37 $\pm$ 0.04  &  2 & SDSS\\
  357       & PSZ2 G083.14$+$66.57 & 14:13:43.73 & $+$43:39:45.0 &  26 &  25 & 0.090 & 0.090 & 0.0893 &  655 $\pm$ 88  &  636 $\pm$ 82  & 0.21 $\pm$ 0.07 & 0.20 $\pm$ 0.07 & 0.19 $\pm$ 0.02  &  1 & SDSS\\
  358       & PSZ2 G083.29$-$31.03 & 22:28:33.72 & $+$20:37:16.3 &  27 &  21 & 0.412 & 0.411 & 0.4116 & 1119 $\pm$ 148 & 1051 $\pm$ 149 & 0.73 $\pm$ 0.25 & 0.63 $\pm$ 0.26 & 0.83 $\pm$ 0.04  &  1 & SDSS\\
  361       & PSZ2 G083.86$+$85.09 & 13:05:45.47 & $+$30:54:25.2 &  25 &  21 & 0.184 & 0.184 &    -   &  693 $\pm$ 95  &  676 $\pm$ 96  & 0.23 $\pm$ 0.08 & 0.22 $\pm$ 0.09 & 0.45 $\pm$ 0.03  &  1 & SDSS\\
  363       & PSZ2 G084.13$-$35.41 & 22:41:58.42 & $+$17:32:23.3 &   7 &  -  & 0.315 &   -   & 0.3020 & 1375 $\pm$ 361 &      -         & 1.62 $\pm$ 1.65 &      -          & 0.51 $\pm$ 0.05  &  2 & SDSS\\
  366       & PSZ2 G084.41$-$12.43 & 21:37:53.37 & $+$35:34:59.8 &  25 &  25 & 0.274 & 0.274 & 0.2763 & 1368 $\pm$ 209 & 1368 $\pm$ 209 & 1.38 $\pm$ 0.50 & 1.38 $\pm$ 0.50 & 0.50 $\pm$ 0.04  &  1 & ITP13\\
\hline 
\end{tabular}
\end{table}
\addtocounter{table}{-1}

\end{landscape}

\begin{landscape}
\begin{table}
\caption{Continued.}
\label{tab:vd4}
\centering
\tiny
\begin{tabular}{c c c c c c c c c c c c c c c c c}
\hline\hline                        
\noalign{\smallskip}
 ID$^1$ & NAME & R.A. & Dec. & N$_{1.5}$ & N$_1$ & z$_{1.5}$ & z$_1$ & z$_{BCG}$ & $\sigma_{200,1.5}$ & $\sigma_{200,1}$ & $M^{dyn}_{500,1.5}$ & $M^{dyn}_{500,1}$ & $M^{SZ}_{500}$ & Flag$^2$ & Data set \\
\noalign{\smallskip}
 &      &      &      &           &       &           &       &           & (km\,s$^{-1}$)     & (km\,s$^{-1}$)   & $(10^{15}\textup{M}_\odot)$  &  $(10^{15}\textup{M}_\odot)$  & $(10^{15}\textup{M}_\odot)$ & & \\
\hline 
\noalign{\smallskip}
  507       & PSZ2 G105.40$-$50.43 & 00:06:21.53 & $+$10:50:49.5 &  30 &  28 & 0.167 & 0.167 &    -   &  709 $\pm$ 90  &  672 $\pm$ 82  & 0.25 $\pm$ 0.08 & 0.22 $\pm$ 0.07 & 0.35 $\pm$ 0.04  &  1 & SDSS\\
  508       & PSZ2 G105.55$+$77.21 & 13:11:08.63 & $+$39:13:36.7 & 109 &  74 & 0.072 & 0.072 & 0.0723 &  843 $\pm$ 64  &  795 $\pm$ 59  & 0.41 $\pm$ 0.05 & 0.35 $\pm$ 0.05 & 0.21 $\pm$ 0.02  &  1 & SDSS\\
  509       & PSZ2 G105.76$+$54.73 & 14:10:18.43 & $+$59:38:03.4 &   9 &   9 & 0.319 & 0.319 & 0.3232 & 1163 $\pm$ 265 & 1134 $\pm$ 253 & 0.97 $\pm$ 0.80 & 0.91 $\pm$ 0.74 & 0.41 $\pm$ 0.04  &  1 & SDSS\\
  515$^{a}$ & PSZ2 G106.41$+$50.82 & 14:25:21.21 & $+$63:09:21.3 &  24 &  24 & 0.140 & 0.140 & 0.1355 & 1564 $\pm$ 212 & 1564 $\pm$ 212 & 2.16 $\pm$ 0.80 & 2.16 $\pm$ 0.80 & 0.27 $\pm$ 0.02  &  0 & SDSS\\
  519       & PSZ2 G107.10$+$65.32 & 13:32:39.39 & $+$50:29:37.7 &  24 &  22 & 0.278 & 0.277 &    -   & 1132 $\pm$ 158 & 1080 $\pm$ 149 & 0.83 $\pm$ 0.31 & 0.73 $\pm$ 0.29 & 0.82 $\pm$ 0.03  &  1 & SDSS\\
  521       & PSZ2 G107.39$-$31.48 & 23:50:36.33 & $+$29:29:29.5 &  17 &  15 & 0.154 & 0.153 & 0.1497 &  657 $\pm$ 108 &  598 $\pm$ 101 & 0.21 $\pm$ 0.10 & 0.17 $\pm$ 0.09 & 0.31 $\pm$ 0.04  &  1 & SDSS\\
\hline 
\end{tabular}
\end{table}
\addtocounter{table}{-1}

\end{landscape}

\begin{landscape}
\begin{table}
\caption{Continued.}
\label{tab:vd5}
\centering
\tiny
\begin{tabular}{c c c c c c c c c c c c c c c c c}
\hline\hline                        
\noalign{\smallskip}
 ID$^1$ & NAME & R.A. & Dec. & N$_{1.5}$ & N$_1$ & z$_{1.5}$ & z$_1$ & z$_{BCG}$ & $\sigma_{200,1.5}$ & $\sigma_{200,1}$ & $M^{dyn}_{500,1.5}$ & $M^{dyn}_{500,1}$ & $M^{SZ}_{500}$ & Flag$^2$ & Data set \\
\noalign{\smallskip}
 &      &      &      &           &       &           &       &           & (km\,s$^{-1}$)     & (km\,s$^{-1}$)   & $(10^{15}\textup{M}_\odot)$  &  $(10^{15}\textup{M}_\odot)$  & $(10^{15}\textup{M}_\odot)$ & & \\
\hline 
\noalign{\smallskip}
  641       & PSZ2 G130.13$-$17.02 & 01:30:55.20 & $+$45:17:18.7 &  20 &  20 & 0.211 & 0.211 & 0.2097 &  908 $\pm$ 139 &  908 $\pm$ 139 & 0.48 $\pm$ 0.21 & 0.48 $\pm$ 0.21 & 0.45 $\pm$ 0.05  &  1 & ITP13\\
  642       & PSZ2 G130.21$-$62.60 & 01:04:55.38 & $+$00:03:36.3 &  24 &  18 & 0.273 & 0.273 & 0.2721 & 1425 $\pm$ 199 & 1330 $\pm$ 204 & 1.54 $\pm$ 0.57 & 1.31 $\pm$ 0.61 & 0.53 $\pm$ 0.05  &  1 & SDSS\\
  644       & PSZ2 G130.64$+$37.16 & 10:47:45.22 & $+$78:00:00.2 &  15 &  15 & 0.472 & 0.472 & 0.4722 & 1166 $\pm$ 197 & 1168 $\pm$ 197 & 0.82 $\pm$ 0.44 & 0.82 $\pm$ 0.44 & 0.35 $\pm$ 0.07  &  1 & LP15\\
  652       & PSZ2 G133.59$+$50.68 & 11:46:48.99 & $+$65:05:01.9 &  57 &  56 & 0.530 & 0.530 & 0.5242 & 1112 $\pm$ 107 & 1171 $\pm$ 100 & 0.65 $\pm$ 0.12 & 0.75 $\pm$ 0.14 & 0.45 $\pm$ 0.05  &  1 & LP15\\
  653       & PSZ2 G133.60$+$69.04 & 12:29:01.82 & $+$47:34:11.3 &  27 &  26 & 0.258 & 0.258 & 0.2588 & 1168 $\pm$ 155 & 1037 $\pm$ 131 & 0.91 $\pm$ 0.31 & 0.66 $\pm$ 0.23 & 0.59 $\pm$ 0.04  &  1 & SDSS\\
  657       & PSZ2 G134.59$+$53.38 & 11:50:49.20 & $+$62:19:48.1 &  17 &  17 & 0.347 & 0.347 & 0.3452 & 1004 $\pm$ 165 &  978 $\pm$ 155 & 0.59 $\pm$ 0.29 & 0.55 $\pm$ 0.27 & 0.38 $\pm$ 0.05  &  1 & SDSS\\
\hline 
\end{tabular}
\end{table}
\addtocounter{table}{-1}

\end{landscape}

\begin{landscape}
\begin{table}
\caption{Continued.}
\label{tab:vd6}
\centering
\tiny
\begin{tabular}{c c c c c c c c c c c c c c c c c}
\hline\hline                        
\noalign{\smallskip}
 ID$^1$ & NAME & R.A. & Dec. & N$_{1.5}$ & N$_1$ & z$_{1.5}$ & z$_1$ & z$_{BCG}$ & $\sigma_{200,1.5}$ & $\sigma_{200,1}$ & $M^{dyn}_{500,1.5}$ & $M^{dyn}_{500,1}$ & $M^{SZ}_{500}$ & Flag$^2$ & Data set \\
\noalign{\smallskip}
 &      &      &      &           &       &           &       &           & (km\,s$^{-1}$)     & (km\,s$^{-1}$)   & $(10^{15}\textup{M}_\odot)$  &  $(10^{15}\textup{M}_\odot)$  & $(10^{15}\textup{M}_\odot)$ & & \\
\hline 
\noalign{\smallskip}
  814       & PSZ2 G171.98$-$40.66 & 03:12:57.83 & $+$08:23:02.0 &  15 &  15 & 0.272 & 0.272 &    -   & 1492 $\pm$ 300 & 1492 $\pm$ 300 & 1.81 $\pm$ 0.98 & 1.81 $\pm$ 0.98 & 1.08 $\pm$ 0.05  &  1 & ITP13\\
  815       & PSZ2 G172.63$+$35.15 & 08:25:29.06 & $+$47:08:00.6 &  50 &  36 & 0.127 & 0.127 & 0.1290 &  733 $\pm$ 73  &  695 $\pm$ 74  & 0.27 $\pm$ 0.06 & 0.24 $\pm$ 0.07 & 0.37 $\pm$ 0.03  &  1 & SDSS\\
  816       & PSZ2 G172.74$+$65.30 & 11:11:43.62 & $+$40:49:14.5 & 101 &  68 & 0.075 & 0.075 & 0.0781 &  633 $\pm$ 50  &  680 $\pm$ 53  & 0.19 $\pm$ 0.02 & 0.23 $\pm$ 0.04 & 0.23 $\pm$ 0.02  &  1 & SDSS\\
  817       & PSZ2 G172.93$+$21.34 & 07:07:37.96 & $+$44:19:22.2 &  17 &  17 & 0.336 & 0.336 & 0.3386 &  990 $\pm$ 160 &  990 $\pm$ 160 & 0.57 $\pm$ 0.28 & 0.57 $\pm$ 0.28 & 0.57 $\pm$ 0.06  &  1 & ITP13\\
  818       & PSZ2 G172.98$-$53.55 & 02:39:53.13 & $-$01:34:56.0 &   7 &   7 & 0.374 & 0.374 & 0.3732 & 1499 $\pm$ 393 & 1462 $\pm$ 377 & 1.97 $\pm$ 2.01 & 1.84 $\pm$ 1.87 & 0.72 $\pm$ 0.05  &  1 & SDSS\\
  821       & PSZ2 G173.90$-$51.89 & 02:45:51.74 & $-$00:42:16.5 &  35 &  32 & 0.181 & 0.181 & 0.1803 &  770 $\pm$ 91  &  779 $\pm$ 88  & 0.31 $\pm$ 0.09 & 0.32 $\pm$ 0.09 & 0.32 $\pm$ 0.04  &  1 & SDSS\\
\hline 
\end{tabular}
\end{table}
\addtocounter{table}{-1}

\end{landscape}

\begin{landscape}
\begin{table}
\caption{Continued.}
\label{tab:vd7}
\centering
\tiny
\begin{tabular}{c c c c c c c c c c c c c c c c c}
\hline\hline                        
\noalign{\smallskip}
 ID$^1$ & NAME & R.A. & Dec. & N$_{1.5}$ & N$_1$ & z$_{1.5}$ & z$_1$ & z$_{BCG}$ & $\sigma_{200,1.5}$ & $\sigma_{200,1}$ & $M^{dyn}_{500,1.5}$ & $M^{dyn}_{500,1}$ & $M^{SZ}_{500}$ & Flag$^2$ & Data set \\
\noalign{\smallskip}
 &      &      &      &           &       &           &       &           & (km\,s$^{-1}$)     & (km\,s$^{-1}$)   & $(10^{15}\textup{M}_\odot)$  &  $(10^{15}\textup{M}_\odot)$  & $(10^{15}\textup{M}_\odot)$ & & \\
\hline 
\noalign{\smallskip}
 1020       & PSZ2 G226.18$+$76.79 & 11:55:15.00 & $+$23:26:40.4 &  29 &  28 & 0.140 & 0.140 &    -   & 1032 $\pm$ 126 & 1036 $\pm$ 126 & 0.70 $\pm$ 0.22 & 0.71 $\pm$ 0.23 & 0.59 $\pm$ 0.02  &  1 & SDSS\\
 1030       & PSZ2 G228.16$+$75.20 & 11:49:41.71 & $+$22:23:57.4 &  13 &  11 & 0.543 & 0.542 &    -   & 1463 $\pm$ 276 & 1501 $\pm$ 300 & 1.46 $\pm$ 0.89 & 1.60 $\pm$ 1.11 & 1.04 $\pm$ 0.05  &  1 & SDSS\\
 1031       & PSZ2 G228.29$+$55.08 & 10:32:50.42 & $+$13:52:42.6 &  11 &   9 & 0.318 & 0.318 & 0.3152 &  763 $\pm$ 157 &  760 $\pm$ 170 & 0.30 $\pm$ 0.21 & 0.31 $\pm$ 0.25 & 0.41 $\pm$ 0.05  &  1 & SDSS\\
 1035       & PSZ2 G228.50$+$34.95 & 09:22:07.65 & $+$03:45:58.8 &   9 &  -  & 0.268 &   -   & 0.2701 &  536 $\pm$ 122 &      -         & 0.12 $\pm$ 0.10 &      -          & 0.10 $\pm$ 0.10  &  0 & SDSS\\
 1036$^{a}$ & PSZ2 G228.62$+$68.44 & 11:23:18.99 & $+$19:38:12.3 &  88 &  74 & 0.103 & 0.103 &    -   & 1352 $\pm$ 109 & 1351 $\pm$ 100 & 1.44 $\pm$ 0.20 & 1.44 $\pm$ 0.23 & 0.27 $\pm$ 0.03  &  0 & SDSS\\
 1038       & PSZ2 G229.37$+$49.70 & 10:14:49.47 & $+$10:39:23.1 &   7 &  -  & 0.195 &   -   & 0.1975 &  932 $\pm$ 245 &      -         & 0.62 $\pm$ 0.63 &      -          & 0.33 $\pm$ 0.04  &  2 & SDSS\\
\end{tabular}

\end{table}
\end{landscape}

\begin{landscape}
\begin{table}
\caption{26 clusters and groups beyond PSZ2.}
\label{tab:vd_others}
\centering
\tiny
\begin{tabular}{c c c c c c c c c c c c c c c c}
\hline\hline                        
\noalign{\smallskip}
Field & R.A. & Dec. & N$_{1.5}$ & N$_1$ & z$_{1.5}$ & z$_1$ & z$_{BCG}$ & $\sigma_{200,1.5}$ & $\sigma_{200,1}$& $M^{dyn}_{500,1.5}$ & $M^{dyn}_{500,1}$ & Data set \\  
\hline 
\noalign{\smallskip}

 PSZ2 G032.12$-$14.96 & 19:43:11.20 & $-$07:24:56.3 &  10 &   8 & 0.378 & 0.378 & 0.3775 &  522 $\pm$ 113 &  535 $\pm$ 128 & 0.11 $\pm$ 0.08 & 0.12 $\pm$ 0.11 & ITP13\\   
 PSZ2 G037.48$+$71.52 & 14:14:34.57 & $+$27:08:35.6 &  17 &  10 & 0.161 & 0.162 &    -   &  621 $\pm$ 102 &  499 $\pm$ 105 & 0.18 $\pm$ 0.09 & 0.11 $\pm$ 0.08 & SDSS\\    
 PSZ2 G039.34$+$73.28 & 14:07:32.82 & $+$27:47:58.3 &  10 &   8 & 0.165 & 0.164 &    -   &  634 $\pm$ 137 &  608 $\pm$ 145 & 0.20 $\pm$ 0.15 & 0.19 $\pm$ 0.17 & SDSS\\    
 PSZ2 G042.54$+$18.02 & 18:02:56.84 & $+$16:21:52.6 &   8 &  -  & 0.258 &   -   &    -   &  414 $\pm$ 101 &      -         & 0.06 $\pm$ 0.06 &      -          & LP15\\    
 PSZ2 G047.48$+$37.37 & 16:50:20.41 & $+$26:58:21.4 &  10 &   8 & 0.230 & 0.230 & 0.2318 &  599 $\pm$ 129 &  494 $\pm$ 118 & 0.17 $\pm$ 0.13 & 0.11 $\pm$ 0.10 & ITP13\\   
 PSZ2 G065.45$+$78.10 & 13:39:00.65 & $+$32:58:12.0 &   7 &  -  & 0.486 &   -   &    -   & 1123 $\pm$ 295 &      -         & 0.84 $\pm$ 0.86 &      -          & SDSS\\    
 PSZ2 G066.59$-$58.51 & 23:07:11.28 & $-$07:31:43.6 &  17 &  17 & 0.334 & 0.334 & 0.3339 &  537 $\pm$ 88  &  523 $\pm$ 83  & 0.11 $\pm$ 0.05 & 0.10 $\pm$ 0.05 & LP15\\
\hline 
\end{tabular}
\end{table}
\end{landscape}

\section{Testing the regression methods}
\label{app:B}

Here we test and validate five different regression methods that account for uncertainties in both axes, for the particular case of the sample discussed in this paper. This study is essential to verify the range of applicability of the methods, and to characterize the existence of statistical biases. 
Noise levels in the data and the intrinsic scatter of the underlying relation play an important role in the recovery of the best-fit estimates. To test these five methods, we perform simulations tailored to mimic the same statistical properties as in our parent sample.
We show that for the noise levels of our reference sample, all the five methods present a bias in some of the recovered parameters. However, all of them are unbiased in the limit of high signal-to-noise (small uncertainties). 

\subsection{Regression methods}
\label{appB1}
We consider the problem of carrying out a linear fit of two variables
with errors in both axes and including intrinsic scatter. We use the following notation. The two variables are given by $x_i$ and $y_i$. Each one of those has measured errors described by Gaussian statistics, with variance $\sigma_{\rm x,i}$ and $\sigma_{\rm y,i}$, respectively. The two variables are tracing underlying quantities $\xi_i$ and $\eta_i$, in such a way that
\begin{equation}
x_i = \xi_i + \epsilon_{x,i}, \qquad
y_i = \eta_i + \epsilon_{y,i}.
\end{equation}
By definition, $\langle \epsilon_{x,i} \rangle = \langle \epsilon_{y,i}\rangle=0$, $\langle \epsilon_{x,i}^2 \rangle=\sigma_{\rm x, i}^2$ and $\langle \epsilon_{y,i}^2 \rangle=\sigma_{\rm y, i}^2$.  
The underlying model that we want to fit for is: 
\begin{equation}
\eta_i = m \xi_i + n + \epsilon_i, 
\end{equation}
with parameters $m$ (slope) and $n$ (intercept). The intrinsic scatter, $\sigma_{\rm int}$, is represented by $\langle \epsilon_{i} \rangle=0$, and $\langle \epsilon_{i}^2 \rangle=\sigma_{\rm int}^2$.
Here we consider the following regression methods:
\begin{enumerate}
\item[i)] The Orthogonal Distance Regression (ODR) method, which uses a modified trust-region Levenberg-Marquardt-type algorithm \citep{ODR} to estimate the function parameters. It is implemented in the python {\it scipy.odr} package.
\item[ii)]  Nukers \citep{Tremaine02} method. It is based on the minimization of the $\chi^2$ function
\begin{equation} \label{eq:chi2}
\chi^2 = \sum_i\frac{(y_i-m x_i -n)^2}{\sigma_{y,i}^2 + m^2\sigma_{x,i}^2}.
\end{equation}
\item[iii)]  Maximum Likelihood Estimator with Uniform prior (MLEU). In this case, the full posterior distribution, assuming Gaussian statistics and flat priors for the three parameters $(m,n,\sigma_{\rm int})$, is given by
\begin{equation} \label{eq:chi2b}
\ln P \propto - \frac{1}{2}\left[\sum_i\frac{(y_i - m x_i - n)^2}{\sigma_{y,i}^2 + m^2\sigma_{x,i}^2 + \sigma_{int}^2} + \ln(\sigma_{y,i}^2 + m^2\sigma_{x,i}^2 + \sigma_{\rm int}^2)\right].
\end{equation}
\item[iv)]  Bivariate Correlated Errors and intrinsic Scatter \citep[BCES,][]{BCES}, which is a Bayesian method commonly used by the galaxy cluster community. We use here the python implementation from {\it astropy.stats} which corresponds to the orthogonal distances method.
\item[v)]  Complete Maximum Likelihood Estimation (CMLE) with correct priors \citep{Kelly07}. Here we use the implementation of this method from \url{ https://github.com/jmeyers314/linmix}.
\end{enumerate}

BCES, MLEU and CMLE methods consider the intrinsic scatter ($\sigma_{\rm int}$) explicitly in their calculations, and indeed both MLEU and CMLE provide an estimation of its value.
The ODR and Nukers methods do not take into account explicitly the intrinsic scatter. However, \citet{Tremaine02} showed how to obtain an estimation of the $\sigma_{\rm int}$ for the Nukers method. Once the best-fit model has been obtained, we evaluate the reduced $\chi^2$ in equation~\ref{eq:chi2}. If this value is smaller than one, then the intrinsic scatter is taken to be zero. Otherwise, the intrinsic scatter is calculated by replacing $\sigma_{y,i}^2$ by $\sigma_{y,i}^2+\sigma_{int}^2$ in the denominator of equation~\ref{eq:chi2} and balancing the right-hand side term until the reduced $\chi^2$ is equal to one. 

\subsection{Simulations}
\label{appB2}

The five methods described in the previous subsection are tested here in their complete forms, fitting simultaneously for the slope, intercept and the intrinsic scatter (if included in the method). In addition, the ODR, Nukers and MLEU are also tested in the particular case of fixing the slope to one, which in our case means that there is no mass dependence of the mass bias. 

To test these methods, we carry out a set of realistic simulations, 
mimicking the sample size (297 objects) and noise conditions that we have in our cluster sample. 
We run three sets of simulations. In the first two sets, we use the same GCs for every iteration while in the last one we generate a set of 297 synthetic clusters for each iteration. In more detail:

\begin{enumerate}
\item Set 1. We use the 297 real clusters from Table \ref{tab:vd}. We assume the estimated SZ masses $M_{500}^{SZ}$ as the true masses $M_{true}$, and we fix the estimated redshift $z$ and the number of cluster members $N_{gal}$ to the real ones. 
\item Set 2. Fitting the properties of the real parent sample from Table~\ref{tab:vd}, we obtain a realistic distribution of dynamical masses, $z$ and $N_{\rm gal}$.  We use these distributions to generate the true simulated mass $M_{\rm true}$, $z$ and $N_{\rm gal}$ of a set of 297 synthetic clusters. 
\item Set 3. We build a set of 297 synthetic clusters in the same way as in 2, but for every iteration.
\end{enumerate}

We use the procedure explained below to obtain the measured SZ and dynamical masses.

Using the \cite{Munari13} relation
\begin{equation} \label{eq:munari2}
\frac{M_{200}^{dyn}}{10^{15}\textup{M}_\odot} = \left(\frac{\sigma_{200}}{A}\right)^\frac{1}{\alpha},
\end{equation}
 we obtain the true velocity dispersion $\sigma_{200}$. The next step is to simulate the measured velocity dispersion which is our observable. Here, for each cluster in every realization we create a set of $N_{gal}$ galaxies which are normally distributed around the $\sigma_{200}$ and we estimate the measured velocity dispersion $\sigma_{measured}$ using these galaxies (only for 2 and 3, the $N_{gal}$ for 1 is fixed). Now, we apply the same procedure as for the read data. We convert the $\sigma_{measured}$ into measured dynamical mass $M_{dyn}$ using Eq.\,\ref{eq:munari}. Then, we correct this mass using the corrections from \cite{Biases} due to the low number of members. The measured uncertainties are directly calculated from Eq.\,C.1 in \cite{Biases}. We do not include the intrinsic scatter of the relation \ref{eq:munari2} because it is less than $5\%$ and the uncertainties in the real data masses are not less than $10\%$ and up to $80\%$ with an average of $40\%$.

On the other hand, we simulate the SZ masses $M_{SZ}$ by calculating the observable $\hat{Y}_{SZ}$ using the inverse procedure than in \planck\ papers. To obtain $\hat{Y}_{SZ}$ we introduce the true simulated masses into the equation:
\begin{equation} \label{eq:Y500}
E^{-\beta}(z)\left[\frac{D^2_A(z)\hat{Y}_{SZ}}{10^{-4}Mpc^2}\right] = Y_*\left[\frac{h}{0.7}\right]^{-2+\alpha}\left[\frac{(1-b)M_{true}}{6\times10^{14}\textup{M}_\odot}\right]^{\alpha},
\end{equation}
where $ D^2_A(z)$ is the angular-diameter distance to redshift $z$ and $E^2(z)=\Omega_m(1+z)^3+\Omega_{\Lambda}$. The coefficients $Y_*$, $\alpha$ and $\beta$ are given in Table 1 in \cite{2016A&A...594A..24P}. Once we have obtained $\hat{Y}_{SZ}$, we include the intrinsic scatter of this relation by adopting a log-normal distribution for the observed $Y_{SZ}$ around its mean value $\hat{Y}_{SZ}$ with $\sigma_{log\ Y}=0.075\pm0.01$ \citep[see][]{2014A&A...571A..20P}. Finally we insert the observed $Y_{SZ}$ into eq. \ref{eq:Y500} to obtain the measured $M_{SZ}$ using the mass bias $(1-b)=0.80$ and we apply a Gaussian random noise based on the real data measured uncertainties.

The theoretically predicted intrinsic scatter in the $M_{dyn}-M_{SZ}$ (Eq.\ \ref{eq:lnBnorm}) is $\sigma_{ln\ M}=0.096$. It is calculated by propagating the intrinsic scatter $\sigma_{log\ Y}$ from the $Y-M$ relation (Eq.\ \ref{eq:Y500}) into eq.\ \ref{eq:lnBnorm}. There are other sources of intrinsic scatter, such as the scatter in eq.\ \ref{eq:munari2}, but in this simulations we only consider $\sigma_{log\ Y}$ as it is the largest.

In our particular case, we assume no dynamical mass bias $(1-b_{dyn})$ so what we are recovering is the SZ bias which is the same as the bias between the SZ and dynamical masses $(1-B)=(1-b)$.

\subsection{Results}
\label{appB3}

\begin{table*}
\tiny
\centering
\caption{Results for the recovered parameters. Input values are $(1-b)=e^n=0.8$, $slope\ (m=1)$, $\sigma_{ln\ M}=0.096$}
\begin{tabular}{c c c c c c c c c c}
\hline \hline
\noalign{\smallskip}

\multirow{2}{*}{Method}      & \multicolumn{3}{c}{$(1-b)=e^n$}                           & \multicolumn{3}{c}{$slope\ (m)$}                          & \multicolumn{3}{c}{$\sigma_{ln\ M}$} \cr

\cline{2-4}
\cline{5-7}
\cline{8-10}
\noalign{\smallskip}
                             &     Set 1     &     Set 2     &     Set 3     &     Set 1     &     Set 2     &     Set 3     &     Set 1     &     Set 2     &     Set 3     \cr

\hline
\noalign{\smallskip}
 \multirow{2}{*}{ODR}        & 0.848 $\pm$ 0.020 & 0.859 $\pm$ 0.021 & 0.861 $\pm$ 0.022 &       1.000       &       1.000       &       1.000       &        $-$        &        $-$        &        $-$        \\
                             & 0.856 $\pm$ 0.027 & 0.859 $\pm$ 0.036 & 0.861 $\pm$ 0.038 & 1.041 $\pm$ 0.056 & 1.003 $\pm$ 0.055 & 1.002 $\pm$ 0.074 &        $-$        &        $-$        &        $-$        \\
                                                                                                                                                     
\hline                                                                                                                                               
\noalign{\smallskip}                                                                                                                                 
 \multirow{2}{*}{Nukers}     & 0.848 $\pm$ 0.020 & 0.859 $\pm$ 0.021 & 0.861 $\pm$ 0.022 &       1.000       &        1.000      &       1.000       & 0.283 $\pm$ 0.067 & 0.276 $\pm$ 0.051 & 0.272 $\pm$ 0.051 \\
                             & 0.857 $\pm$ 0.056 & 0.859 $\pm$ 0.032 & 0.862 $\pm$ 0.036 & 1.042 $\pm$ 0.056 & 1.005 $\pm$ 0.051 & 1.003 $\pm$ 0.059 & 0.308 $\pm$ 0.079 & 0.280 $\pm$ 0.057 & 0.276 $\pm$ 0.060 \\
                                                                                                                                                                                      
\hline                                                                                                                                                                                
\noalign{\smallskip}                                                                                                                                                                  
 \multirow{2}{*}{MLE }       & 0.868 $\pm$ 0.022 & 0.875 $\pm$ 0.023 & 0.876 $\pm$ 0.024 &       1.000       &        1.000      &       1.000       & 0.222 $\pm$ 0.043 & 0.241 $\pm$ 0.039 & 0.239 $\pm$ 0.041 \\
                             & 0.772 $\pm$ 0.010 & 0.677 $\pm$ 0.015 & 0.672 $\pm$ 0.020 & 0.404 $\pm$ 0.055 & 0.522 $\pm$ 0.040 & 0.501 $\pm$ 0.047 & 0.308 $\pm$ 0.027 & 0.285 $\pm$ 0.028 & 0.280 $\pm$ 0.027 \\
                                                                                                                                                     
\hline                                                                                                                                               
\noalign{\smallskip}                                                                                                                                 
 BCES                        & 0.852 $\pm$ 0.033 & 0.808 $\pm$ 0.064 & 0.800 $\pm$ 0.069 & 0.800 $\pm$ 0.146 & 0.848 $\pm$ 0.147 & 0.836 $\pm$ 0.159 & 0.307 $\pm$ 0.056 & 0.253 $\pm$ 0.066 &  0.260 $\pm$ 0.090    \\
                                                                                                                                                     
\hline                                                                                                                                               
\noalign{\smallskip}                                                                                                                                 
 CMLE                        & 0.846 $\pm$ 0.021 & 0.794 $\pm$ 0.027 & 0.786 $\pm$ 0.028 & 0.830 $\pm$ 0.086 & 0.808 $\pm$ 0.060 & 0.792 $\pm$ 0.063 & 0.043 $\pm$ 0.016 & 0.052 $\pm$ 0.014 & 0.050 $\pm$ 0.014 \\

\hline
\noalign{\smallskip}

\end{tabular}
\label{tab:regtest}

\end{table*}

Table \ref{tab:regtest} shows the results of the regression tests performed over the simulations described in appendix \ref{appB2}. The first column names the regression method. The second column presents the median value of the mass bias $(1-b)$. The third column shows the median value for the slope $m$. The last column presents the natural logarithm of the intrinsic scatter when available.

\begin{figure}
\centering \includegraphics[width=\columnwidth]{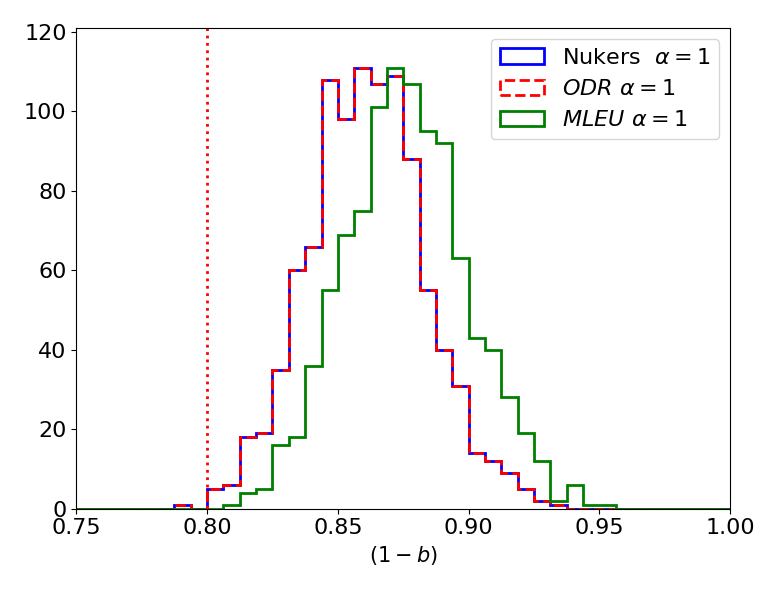}
\caption{Distribution of the estimation of the mass bias parameter when fixing the slope to one for Nukers, ODR and MLEU methods (Set 3). Vertical dashed line represents the input value $(1-b)=0.8$.}
\label{fig:ApBfix}
\end{figure}

First, we discuss the results for the case of no mass dependence in the mass bias $(m=1)$. There is no particular method that recovers the $(1-b)=0.8$. All three tested methods are biased, regardless of the initial settings of the simulations. As shown in figure \ref{fig:ApBfix} the ODR and the Nukers are biased upwards by a $7\%$ while the MLEU is biased by a $9\%$ in the same direction. We consider this effect as a true bias as the standard deviations in the three methods are not greater than $3\%$, see figure \ref{fig:ApBfix} and table \ref{tab:regtest}. The MLEU estimates the intrinsic scatter of the relation as $\sigma_{ln\ M}=0.24\pm0.04$ which is more than 3-$\sigma$ away from the predicted one $\sigma_{ln\ M}=0.096$. The estimation that comes out from the Nukers method is $\sigma_{ln\ M}=0.27\pm0.05$, also more than 3-$\sigma$ away. This might be the reason why the methods are biased, the overestimation of the intrinsic scatter may lead into a biased estimation of the intercept. The explanation for the latter is that the methods may not be able to disentangle the difference between the intrinsic scatter and the measurement errors as they are, on average, four times larger. Another possible explanation is that the error propagation might not be as precise as required. We use symmetric errors in the logarithm space and they are calculated as the uncertainty over the quantity in the real space. This is just an approximation that with our big uncertainties might produce this type of bias.

We also perform the same analysis varying the input value of the mass bias from 0.6 to 1.2 obtaining the same results as explained above. In every case, the $(1-b)$ parameter and the slope are biased in the same percentage as when using $(1-b)=0.8$.

Now, we discuss the case of a possible dependence of the mass bias parameter with the mass, in other words, letting the slope of the regression free to vary. The following results are independent of the initial settings of the simulation as shown in table \ref{tab:regtest}. The ODR and the Nukers methods, as in the case of fixed slope, are biased in the recovery of the parameter $(1-b)$ exactly in the same percentage as the recovery of the slope is almost perfect. The only difference is that the standard deviation is greater for obvious reasons. The MLEU, the BCES and the CMLE fail completely when trying to recover the slope. Although the BCES and the CMLE do recover the mass bias parameter, these methods must not be trusted as the slope they recover is between $15$ and $20\%$ lower than the input value. The MLEU fails catastrophically in both tasks. The bad behavior of these methods might be caused by the wrong estimation of the intrinsic scatter because of the confusion with the huge measurement errors, similar to the case of fixed slope.

\begin{figure}
\centering \includegraphics[width=\columnwidth]{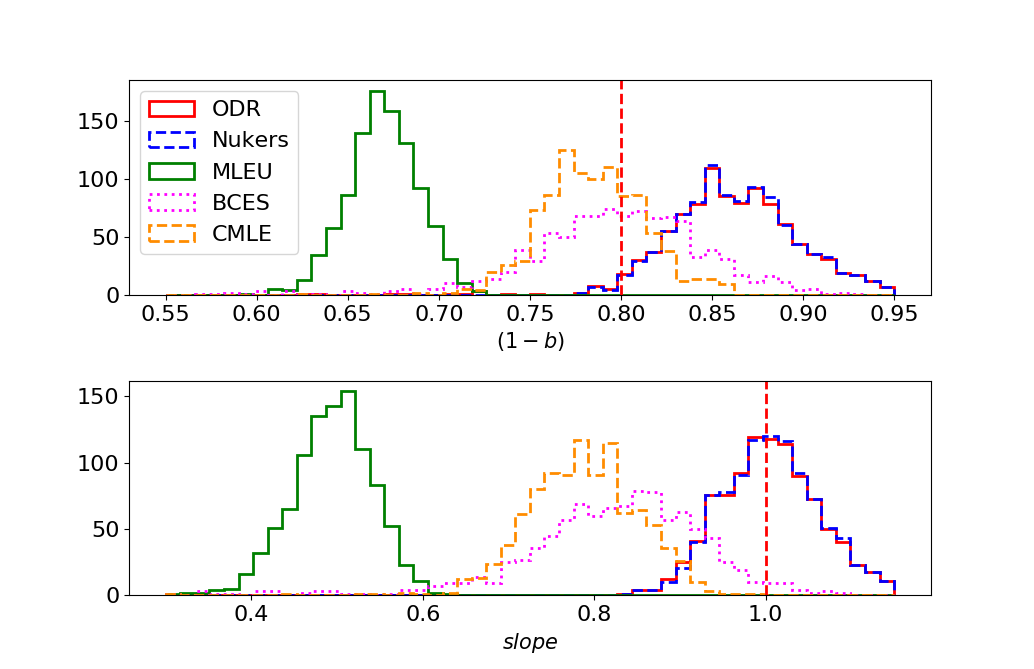}
\caption{Distribution of the estimation of the mass bias parameter (top panel) and the slope (bottom panel) for the five tested regression methods (Set 3). Vertical dashed lines represent the input values of the simulation $(1-b)=0.8$ and $\alpha=1$.}
\label{fig:ApB}
\end{figure}

\begin{figure}
\centering \includegraphics[width=\columnwidth]{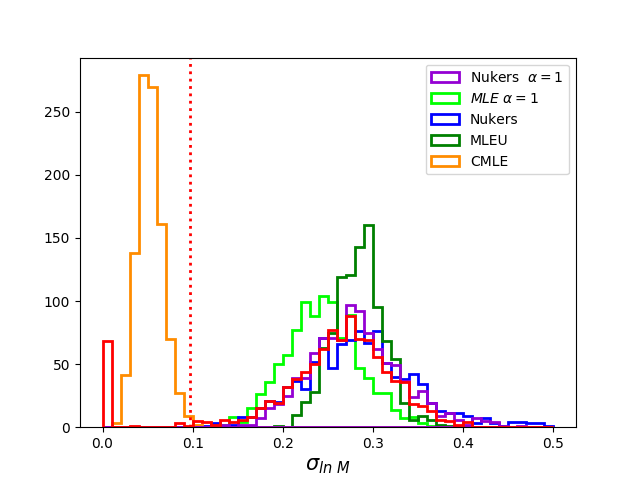}
\caption{Distribution of the estimation of the intrinsic scatter for Nukers, MLEU, BCES and CMLE (Set 3). Vertical dashed line represents the theoretical value of the intrinsic scatter $\sigma_{\ln M}=0.096$.}
\label{fig:ApBsig}
\end{figure}

Figure~\ref{fig:ApBsig} shows the distribution of the intrinsic scatter for the three methods that estimate it. We think this is one of the key questions and why the methods do not recover properly the input values of the parameters. As explained in the previous section, the theoretical intrinsic scatter can be calculated and there is no single method that estimate it correctly. We perform the same simulations setting the intrinsic scatter to zero and we find very similar results to those discussed above. We also perform the simulations setting the measurement errors two orders of magnitude lower. In this case every method recovers properly the input values even the intrinsic scatter within 1-, 2-$\sigma$ depending on the method.

In addition, we have also explored if a binning approach improves the result. For this study, we divided the simulated sample (set 3) in 5, 10 and 15 bins. In general, the binning approach increases the errors on the recovered parameters, independently of the number of bins used, being still fully consistent with the case of no binning. It slightly increases the bias in the recovery of the 1-b from 6\% to 10\%, but still consistent with the value for no binning. We conclude that this binning approach will not improve the results so we do not use it in this paper.

We conclude that there is no correct regression method to use in this configuration, in other words, each method is either biased or gives wrong results. The main source of trouble are the big measurements errors combined with the intrinsic scatter. We select the Nukers method as our reference method for two main reasons. It gives a robust estimation of the slope when we let it vary, and it has a small bias in the $(1-b)$ parameter which we can correct or account for. Other methods might give better estimation of the $(1-b)$ parameter, like the MLEU which is less biased than the Nukers but this method does not recovered correctly the slope (see Fig.~\ref{fig:ApB}). In combination of $(1-b)$ and slope we recommend to use the Nukers method for our particular set of data.

\end{document}